\documentstyle[12pt,aaspp4,epsf]{article}		% ApJ and AJ preprint
\slugcomment{2001, ApJ, 548, 20 Feb}

\begin{document}

\title{Abundances and Evolution of Lithium in the Galactic Halo and 
Disk\footnote{Based on observations obtained with 
the University College London \'echelle spectrograph (UCLES) on the
Anglo-Australian Telescope (AAT) and the Utrecht \'echelle spectrograph (UES)
on the William Herschel Telescope (WHT).}}

\author{Sean G. Ryan}
\affil{Dept of Physics and Astronomy, The Open University, Walton Hall, 
MK7 6AA, UK.
email: s.g.ryan@open.ac.uk}

\author{Toshitaka Kajino}
\affil{National Astronomical Observatory of Japan, Osawa 2-21-1, Mitaka, 
Tokyo 181-8588, Japan, and
Department of Astronomy, University of Tokyo, 7-3-1 Hongo, Bunkyo-ku,
Tokyo 113-0015, Japan.
email: kajino@ferio.mtk.nao.ac.jp}

\author{Timothy C. Beers}
\affil{Dept of Physics and Astronomy, Michigan State University, East Lansing,
MI 48824, USA. email: beers@pa.msu.edu}

\author{Takeru Ken Suzuki}
\affil{National Astronomical Observatory of Japan, Osawa 2-21-1, Mitaka, 
Tokyo 181-8588, Japan, and
Department of Astronomy, University of Tokyo, 7-3-1 Hongo, Bunkyo-ku,
Tokyo 113-0015, Japan.
email: stakeru@th.nao.ac.jp}

\author{Donatella Romano}
\affil{SISSA/ISAS, Via Beirut, 2-4, 34014, Trieste, Italy. email: romano@sissa.it}

\author{Francesca Matteucci}
\affil{Dipartimento di Astronomia, Universit\`a di Trieste, Via G. B. Tiepolo 11, 34131 Trieste, Italy,  and SISSA/ISAS, Via Beirut, 2-4, 34014, Trieste, Italy. email: matteucci@ts.astro.it}

\and

\author{Katarina Rosolankova}
\affil{Dept of Physics and Astronomy, The Open University, Walton Hall, 
Milton Keynes MK7 6AA, UK.
email: katarina.rosolankova@st-hildas.oxford.ac.uk}

\begin{abstract}

We have measured the Li abundance of 18 stars with
$-2~^<_\sim$~[Fe/H]~$^<_\sim~-1$ and 6000~K~$^<_\sim~T_{\rm
eff}~^<_\sim$~6400~K, a parameter range that was poorly represented in previous
studies.  We examine the Galactic chemical evolution (GCE) of this element,
combining these data with previous samples of turnoff stars over the full range
of halo metallicities.  We find that $A$(Li) increases from a level of $~\sim
2.10$ at [Fe/H]~=~$-3.5$, to $\sim 2.40$ at [Fe/H]~=~$-1.0$, where $A$(Li)~=
log$_{10}$($n$(Li)/$n$(H))~+~12.00.

We compare the observations with several GCE calculations, including existing
one-zone models, and a new model developed in the framework of inhomogeneous
evolution of the Galactic halo.  We show that Li evolved at a constant rate
relative to iron throughout the halo and old-disk epochs, but that during the
formation of young-disk stars, the production of Li relative to iron increased
significantly.  These observations can be understood in the context of models
in which post-primordial Li evolution during the halo and old-disk epochs is
dominated by Galactic cosmic ray fusion and spallation reactions, with some
contribution from the $\nu$-process in supernovae.  The onset of more efficient
Li production (relative to iron) in the young disk coincides with the
appearance of Li from novae and AGB stars.  The major challenge facing the
models is to reconcile the mild evolution of Li during the halo and old-disk
phases with the more efficient production (relative to iron) at
[Fe/H]~$>~-0.5$. We speculate that cool-bottom processing (production) of Li in
low-mass stars may provide an important late-appearing source of Li, without
attendant Fe production, that might explain the Li production in the young
disk.
\end{abstract}

\keywords{
early Universe
---
cosmology: observations
---
nuclear reactions, nucleosynthesis, abundances 
--- 
stars: abundances 
---  
stars: Population II 
---
Galaxy: halo
}

\vfill
\eject

\section{Introduction}					% 1

Lithium plays several valuable roles as a diagnostic of stellar and Galactic
evolution.  As the only metal synthesized in significant quantities in the big
bang, $^7$Li provides a rare constraint on the baryon density of the universe
(e.g., Ryan et al. 2000).  As an element destroyed in stars where the
temperature exceeds $2.5\times 10^6$~K, its survival at the stellar surface
indicates the degree of exchange of material between the surface and interior
via convection, diffusion and other processes.  Thirdly, as a product of
spallation and fusion reactions and of stellar sources, it provides a measure
of the chemical evolution of the Galaxy.

In practice, the three roles of Li cannot be treated in isolation.  The
primordial (big bang) abundance cannot be determined without knowing the
sources and sinks of the element, and the degree of mixing for stars of
different metallicity cannot be determined from the observations unless the
contribution of Galactic production is known.  Normally, one attempts to reduce
the complexity of the problem by isolating one or two parts.  For example, the
near-constancy of the $^7$Li abundance in warm halo stars over a range of
effective temperature and metallicity led Spite \& Spite (1982) to conclude
that destruction of Li in those stars and its production in the course of
Galactic chemical evolution were negligible. As a result, they argued that
it was proper to consider the observed Li abundance as the primordial one,
hardly altered.  The view that Li in these objects was unaltered was supported
empirically by the small spread in Li abundances and by classical stellar
evolution models (e.g., Yale `standard' models [Deliyannis, Demarque, \& Kawaler
1990]) which showed negligible levels of pre-main sequence and main sequence
destruction even over the long ($\simeq$13 Gyr) lives of the objects.  (This
contrasted with the considerable destruction seen in some young open clusters
[e.g., Hobbs \& Pilachowski 1988], but is understood as the depth of the surface
convection zone being less in stars of lower metallicity, thus not reaching the
depths required for burning at $T \ge 2.5\times 10^6$~K).

Challenges to the Li survival hypothesis have come from both theoretical and
observational sources.  More complex (and hopefully more realistic, but also
more uncertain) stellar evolution models, involving rotationally-induced
mixing, were found to be able to deplete significant fractions of the Li in
these objects.  Early models suggested as much as 90\% could be lost
(Pinsonneault, Deliyannis, \& Demarque 1992), though later work suggested
perhaps half might be destroyed (Pinsonneault et al. 1999).  (The downward
revision of the figure was driven partly by the models and partly by
observational data.) Coupled to this theoretical work were claims of
significant scatter in the observed abundances, inconsistent with a single
primordial value (Deliyannis, Pinsonneault, \& Duncan 1993; Thorburn 1994).
The most recent observations show, however, that the intrinsic scatter in a
sample of 22 halo field turnoff stars is $\sigma_{\rm int}~<~0.02$~dex, and
does not support the proposition of more than 0.1 dex $^7$Li depletion by the
rotational mixing mechanism (Ryan, Norris, \& Beers 1999). Although this field
star sample places very tight limits on the intrinsic spread in $^7$Li, there
is evidence of at least some star-to-star differences in the halo.  Boesgaard
et al. (1998) find a spread in $^7$Li abundance amongst subgiants in M92 (see
Figure 1), Deliyannis (1999, priv.comm.) finds $^7$Li differences between the
extremely metal-poor field stars G64-12 and G64-37, stars which otherwise
appear very similar to one another.  Furthermore, there exist a small number of
very Li-deficient stars which are otherwise indistinguishable from halo Li
``preservers'' (Hobbs, Welty, \& Thorburn 1991; Thorburn 1992; Spite et al.
1993; Norris et al. 1997; Ryan, Norris, \& Beers 1998).  However, the different
(dense) stellar environment of the M92 stars and small volume of the Galaxy it
samples, and the rarity of the other cases, leads us to view the small observed
spread of Li abundance in the recent halo-star sample ($\sigma_{\rm
int}~<~0.02$) as representative of the majority of the halo.

In addition to discussions of the intrinsically thin Spite halo Li plateau,
claims have been made of the existence of dependencies of observed Li abundance
upon $T_{\rm eff}$ and [Fe/H].  Thorburn (1994), Norris, Ryan, \& Stringfellow
(1994), and Ryan et al.  (1996a) all found the halo observations to require
significantly non-zero coefficients to fits of the form $A{\rm
(Li)}~=~A_0~+~A_1{{T_{\rm eff}}\over{100~{\rm K}}}~+~A_2{\rm [Fe/H]}$, where
$A$(Li)~= log$_{10}$($n$(Li)/$n$(H))~+~12.00.  Typical estimated values of the
coefficients were $A_1$~=~0.03 and $A_2$~=~0.14.  The coefficient on $T_{\rm
eff}$, $A_1$, may depend crucially on the adopted temperature scale.  The
optical photometric scales used in the cited studies were challenged by
Bonifacio \& Molaro (1997) who used temperatures derived from application of
the infrared flux method (IRFM), whereupon they concluded that both $A_1$ and
$A_2$ were consistent with zero, which is to say that the Spite plateau is
flat.  The IRFM scale has often been proposed as less likely to be affected by
metallicity-dependent systematic errors (Saxner \& Hammarb\"ack 1985; Magain
1987), but the uncertainties in the $T_{\rm eff}$ of any individual star are
still considerable, with $\sigma_{T_{\rm eff}}~\simeq~100$~K (Alonso, Arribas,
\& Mart\'inez--Roger 1996; Bonifacio \& Molaro 1997).  The other photometric
scales can at least lead to small internal errors, $\sigma_{T{\rm
eff}}~\simeq~30-40$~K (Ryan et al. 1999), and a mean error of 55~K in the 
present sample, but possibly with less reliable
external systematics.  The trade-off is that the IRFM may deliver better
systematics but at the expense of introducing greater internal scatter.
Hopefully, improvements in the systematics of the optical temperature scales
and in the internal errors of the IRFM scale will be achieved, and we will be
able to clarify the size of the $A_1$ term in the near future.

Bonifacio \& Molaro (1997) also found the $A_2$ term (metallicity coefficient)
to be consistent with zero, but the work by Ryan et al. (1999), which achieved
errors as small as $A$(Li)~=~0.03 dex for most stars, again found a significant
value: $A_2~=~0.118~\pm~0.023$.  Ryan et al. traced the main difference between
these two results to some substantial differences in the stellar metallicities
adopted (from the literature) by the two studies.  After comparing with a third
[Fe/H] estimate (the homogeneously applied estimator of metallicity obtained by
Beers et al. 1999), they argued that the [Fe/H] values adopted by Bonifacio \&
Molaro had sufficiently large errors to obscure the $A_2$ analysis.  Ryan et
al. further argued that measurements of $^6$Li in HD~84937 and
BD~+26$^\circ$2578 (Smith, Lambert, \& Nissen 1993, 1998; Cayrel et al. 1999;
Hobbs \& Thorburn 1994, 1997), and now also in HD~140283 (Deliyannis \& Ryan
2000), likewise evidenced the contribution of Galactic chemical evolution (GCE)
to Li production even in stars with [Fe/H]~$\sim~-2.5$, since $^6$Li is thought
to be exclusively post-primordial.  The contributions to the total Li
inferred from $A_2$ and from $^6$Li were found to be compatible.  From
application of the inhomogeneous GCE model of Suzuki, Yoshii, \& Kajino (1999),
Suzuki, Yoshii, \& Beers (2000a) argue that a non-zero slope in the
relationship between $A$(Li) and [Fe/H] (of the same order of magnitude as that
observed) must arise in the early Galaxy from Li production associated with the
spallation reactions that give rise to Be and B.

In their earlier study, Ryan et al. (1996a) noted that an observational bias 
existed in the available Li data for halo stars.  In the quest for the
primordial lithium abundance, observers had studied progressively more
metal-poor stars, but had examined rather fewer at [Fe/H]~$\sim~-1.5$.
Moreover, those that were examined at higher [Fe/H] were invariably cooler than
the more metal-poor ones, potentially complicating the analysis of the
coefficients $A_0$, $A_1$, and $A_2$ by the inadvertent introduction of
collinearity in the predictor variables.  To address both of these
difficulties, we set out to measure a sample of hotter (6000~K~$^<_\sim~T_{\rm
eff}~^<_\sim$~6400~K), more metal-rich ($-2~^<_\sim$~[Fe/H]~$^<_\sim~-1$)
stars.  The sample selection, observations, and abundance analysis are
discussed in the following sections.  We then combine the new data on 14 stars,
which correct the previous paucity of warmer, higher metallicity halo stars
with existing observations, and examine GCE of Li.  In addition to these 14
stars we report on four stars with $A$(Li)~$<~1.7$, two of which are newly
discovered extremely Li-deficient halo stars.  These exceptional stars are
discussed in detail in a separate paper (Ryan et al. 2001).

\section{Sample Selection, Observations and Abundance Analysis} %2

\subsection{Sample selection and observation}

We sought to address two problems:
\newline
(1) the lack of Li measurements in stars with 
$-2~^<_\sim$~[Fe/H]~$^<_\sim~-1$, with which 
to study GCE of this element, and
\newline
(2) the selection bias against warm stars with 
6000~K~$^<_\sim~T_{\rm eff}~^<_\sim$~6400~K
in this metallicity interval.

We searched the catalogues of Schuster \& Nissen (1988, 1989), Schuster,
Parrao, \& Contreras Martinez (1993), Ryan (1989), Ryan \& Norris (1991), and
Carney et al. (1994) for stars in these [Fe/H] and $T_{\rm eff}$ ranges.
Observations were obtained with the University College London \'echelle
spectrograph (UCLES) on the Anglo-Australian Telescope on 1996 September 24 and
with the Utrecht \'echelle spectrograph (UES) on the William Herschel Telescope
(WHT) on 1997 August 23.  Both spectrographs, which have almost identical
configurations, were set up to deliver $\lambda/\Delta\lambda~\simeq~50000$ at
the Li~6707~\AA\ doublet.  Both observing runs utilised 79~lines~mm$^{-1}$
gratings, which allow a slit length of 14~arcsec to ensure adequate sampling of
the background (sky and scattered light) contribution.

The stars for which spectra were obtained are listed in Table 1.  The first
column lists the star name(s).  Columns (2) and (3) list the 2000.0 epoch
positions.  Apparent magnitudes and colors, as well as estimates of the
reddening in the direction to each star, taken from the references listed
above, are listed in columns (4)-(9).\footnote{Although no pre-selection on Li
abundance was made in the assemblage of our sample, one unexpected result was
the inclusion of four Li-deficient stars, two of which were new discoveries.
Table~1 is therefore separated into two parts in recognition of this.} The
[Fe/H] measurements provided in column (10), also taken from the references
above, are based on either medium-resolution spectra or photometric estimates
that, over this metallicity range, are expected to be accurate to $\sigma_{\rm
[Fe/H]}~=~0.15$--0.20~dex (Schuster \& Nissen 1989; Ryan \& Norris 1991; Carney
et al. 1994).  Five of the stars in Table 1 have independent estimates of
[Fe/H] derived from medium-resolution spectroscopy reported by Beers et al.
(1999).  The mean offset is [Fe/H]$_{\rm AK2} - {\rm [Fe/H]}_{\rm lit} =
-0.08$, with an RMS scatter of 0.13 dex, which provides additional evidence
that the metallicities used herein are secure.

Errors in [Fe/H] will affect the Li abundances derived below in three ways. 
Firstly, they
will cause a model atmosphere of the wrong metallicity to be used. The impact
of this is completely negligible, as Table 5 of Ryan et al. (1996) shows.  A
star with $T_{\rm eff}$ = 6300~K, [Fe/H] $\simeq$ $-1.5$, and $A$(Li) = 2.20
would give rise to an error in $A$(Li) of only 0.002~dex for a 0.2~dex error in
[Fe/H].  
Secondly, an error in [Fe/H] would also cause an
incorrect effective temperature to be adopted. 
In the $b-y$ calibrations of Magain (1987), a metallicity error of 0.2~dex at
[Fe/H] = $-1$ and $T_{\rm eff}$ = 6300~K would induce a temperature error of
12~K, which corresponds to only 0.010~dex in $A$(Li). 
This error must of course be added to those arising from the other sources.
Thirdly, an error in
[Fe/H] would cause a star to be shifted along the {\it x}-axis in a $A$(Li) vs
[Fe/H] diagram; the impact of the error in that case depends on the model to
which the data are being compared, and can be assessed from the error bars
shown in such a figure.

\subsection{Effective temperatures and uncertainties}

Effective temperatures were calculated using the same B--V, R--I$_{\rm C}$, and
$b-y$ calibrations as in Ryan et al. (1996a) to maintain consistency with that
work, but with the addition of the V--R$_{\rm C}$ calibration of Bell \& Oke
(1986).  
Ryan et al. found
``very good agreement between B--V and $b-y$ temperatures'', but reported that
``the R--I temperatures exceed the B--V temperatures on average by perhaps 50~K
for the cooler half of the sample, but the systematics are too marginal to 
justify adjusting the scales further.'' As we will combine our present sample
with that of Ryan et al., to assess the impact of 
including hotter, more metal-rich stars,
we utilise the same procedure.

In contrast to the result for the broad 1996 sample where systematic
differences were ``too marginal to 
justify adjusting the scales'', Ryan et al. (1999) found some well defined
and larger offsets for the Bell \& Oke (1986) scales in a narrowly defined 
subset of
very metal-poor ([Fe/H] $\la$ -2.2) and hotter ($T_{\rm eff} \ga 6000$~K) stars.
For this narrowly defined subset of stars, offsets were made to the 
Bell \& Oke scales for those very metal-poor stars of up to 165~K.
Because such offsets were not discernable for the broad (1996) sample, one may
be concerned about the impact of unadjusted systematic errors that remain
embedded in the Ryan et al. (1996) temperatures. For this reason, we
adopted a more conservative approach in this work compared with Ryan et al.
(1996) for computing the {\it uncertainties} on $T_{\rm eff}$. 

To recap, Ryan et al. (1996) propagated errors in each individual photometric 
index and the reddening estimates, and combined these on the assumption that 
they fully captured the error sources. A more conservative approach would have 
been to take the greater of this value or the index-to-index standard deviation 
where two or more colors were available. Indeed, we adopt this more conservative
approach in the current work. As we inspect the index-to-index scatter, our
revised approach will also be inflated by any imbedded systematic differences
between different temperature scales used in the $T_{\rm eff}$ calculation.
However, we find that the more conservative 
approach makes very little difference quantitatively. The uncertainties for the 
current work ranged up to 130~K, with a mean value of 55~K. To be especially 
conservative, we assigned this mean estimate to stars with only a single color
and to those stars whose formal estimate was (probably fortuitously) less than 
55~K. That is, an error in $T_{\rm eff}$ of 55~K is the {\it smallest} we claim
in the current work. In comparison, the errors reported in the 1996 study
ranged from 32~K to 180~K, with a mean of 52~K. The resulting difference between
approaches has almost no effect on the claimed mean error or even the range
of errors deduced. Had we considered the index-to-index scatter in 1996, the
mean error would have been only marginally increased, to 62~K.

As we will combine the present sample and the 1996 one, how should we
regard the 1996 error estimates in hindsight? While a more conservative
error-estimation procedure could have been adopted, and some individual stars
would certainly have had different values quoted, it appears that both the
mean error and the range of errors quoted would not have been significantly
different. Note also that these more conservative estimates are sensitive
to embedded systematic differences between $T_{\rm eff}$ scales for different
photometric indices, since these would increase the star-to-star scatter, yet
the conservatively computed errors for the sample are similar in both range and
mean. For this reason, we have chosen to adopt the published 1996 values unchanged.

As a final comment, we reemphasise that the discussion of errors here has
centred on relative (star-to-star) differences.
It is clear from the various color vs
effective temperature transformations discussed that the adopted zero-point could
still be in error by 100~K or more.
However, zeropoint errors will affect most stars similarly,
and hence will result in an overall translation of observation data moreso
than a rearrangement of the data relative to one another.

\subsection{Spectral analysis}

The spectra were reduced in IRAF using conventional techniques, and final data
are shown in Figure 2.  Table~1 records the telescope used, and the S/N per
50~m\AA\ pixel (taken as the lesser of the S/N based on photon statistics and
the measured scatter in the continuum) in columns (13) and (14), respectively.
Equivalent widths of the Li lines, $W$, were measured for the stars in each of
two ways.  First, a Gaussian fit was made to the Li lines and the equivalent
width and FWHM were recorded.  Once all stars were measured, the mean FWHM of
the fitted Gaussians was computed for each spectrograph.  A second series of
measurements was then made, with the Gaussian FWHM for each spectrograph fixed
at the mean value.  This is done because the $^7$Li doublet, being broader than
the instrumental resolution, is not expected to vary in FWHM from star to star
since all have ``weak'' ($W/\lambda ~<~10^{-6}$) Li lines.  Finally, the two
equivalent width measurements were averaged, and are listed in column (15) of
Table~1. The error in the measured Li equivalent width, reported in column
(16), is based on the error model $\sigma_W~=~{{184}\over{S/N_{50}}}$ (Ryan et
al.  1999).

To maintain consistency with the analysis of Ryan et al. (1996a; also 1999),
the same computations of $A$(Li) from $T_{\rm eff}$, $W$, and [Fe/H] were used.
To recap, these utilised Bell (1983, private communication) stellar atmosphere
models and the spectrum synthesis code of Cottrell \& Norris (1978) to compute
the $^7$Li doublet using four components for the fine structure and hyperfine
structure.  The inferred abundances and their errors, given as the quadratic
sum of separate terms for $\sigma_W$ and $\sigma_T$, are listed in the final
columns of Table 1. 

Figure 3 shows the distribution of the stars in our present analysis in the
$T_{\rm eff}$, [Fe/H] plane.  The new observations are seen to substantially
correct the previous deficit of warmer, more metal-rich systems, though
additional observations of similar stars are certainly warranted.

Radial velocities measured from our spectra are given in Table~2 (in
km~s$^{-1}$). The {\it internal} error estimates ($\sigma_v$, tabulated) are
only 0.1--0.3~km~s$^{-1}$, but similar work by us in the past has suggested an
{\it external} error of 0.3~km~s$^{-1}$ (Ryan et al. 1999). Previous
measurements from \'echelle observations by Carney et al. (1994), which are
accurate to $\sim 0.1$~km~s$^{-1}$, or from medium-resolution spectra of Ryan
\& Norris (1991), accurate to only $\sigma_v~=~7$~km~s$^{-1}$, are included for
comparison.  Stars already identified as spectroscopic binaries are explicitly
noted.  The mean difference $\langle v_{\rm rad} - v_{\rm
CLLA94}\rangle$~=~$-0.35$~km~s$^{-1}$, with an RMS difference of
0.68~km~s$^{-1}$. This is larger than the 0.3~km~s$^{-1}$ accuracy expected,
and may indicate the presence of unidentified low-amplitude and/or long period
binaries in the sample. However, the largest absolute-value residual against
the Carney et al. measurements is only 1.4~km~s$^{-1}$, making it difficult to
distinguish the remaining binaries with certainty. Residuals between the new
data and the measurements of Ryan \& Norris are consistent with the velocity
errors arising from the lower resolution of that older data.

Any concern whether the stars genuinely belong to the halo rather than the disk
populations (since metallicity alone is a poor discriminant at intermediate
abundances) can be dispelled by consideration of the $U, V, W$ space
velocities.  Those shown in columns (6)-(8) of Table~2 are heliocentric
velocities from Ryan \& Norris (1991), or local standard of rest (LSR)
velocities from Carney et al.  (1994) if the former are not available.  The
sole exception is BD$-31^\circ$305, which had not been studied in those works
and was computed here following the precepts of Ryan \& Norris.  The Ryan \&
Norris velocities are used in preference because their distance scale shows
better agreement with Hipparcos measurements for the program stars (see
Figure~4); the Carney et al.  distances tend to underestimate those from
Hipparcos.  All except HD~16031 have velocities in excess of 100~km~s$^{-1}$,
some substantially so, removing any doubt that they are correctly associated
with the halo. For completeness, we also tabulate absolute V magnitudes based
on Hipparcos parallaxes (and errors) in column (2) of the table.

\section{The Li Abundances} %3

Figure 5 shows the newly obtained Li abundances, along with the literature
sample, for the abundance interval $-2.2~<~$[Fe/H]~$<~-1.0$, representing an
extension of the sample discussed by Ryan et al. (1996a) in their Fig. 3(b).
(We restrict the sample to a $\sim$~1~dex interval of [Fe/H] because any
metallicity-dependence would increase the spread in this diagram.) The new
observations are consistent with the older data, and show that the results of
the 1996 analysis, and the claim of the existence of a significant dependence
of $A$(Li) on T$_{\rm eff}$, were not caused by the inherited selection bias
against more metal-rich turnoff stars illustrated in Figure 3.  We quantify
this statement below.

Having added data to the high metallicity regime, and also now having the
benefit of the improved observations of Ryan et al. (1999), it is important to
ask whether the correlations between $A$(Li), T$_{\rm eff}$, and [Fe/H] found
by Ryan et al. (1996a) are still present.  As the current paper is primarily a
study of GCE, we only present an abbreviated discussion of the correlations
here, and refer the reader interested in a more detailed statistical analysis
to Appendix 1.

We have performed bivariate fits of the form $A$(Li)~=~$A_0$ + $A_1{{T_{\rm
eff}}\over{100}}$ + $A_2$[Fe/H] to the stellar sample assembled in Ryan et al.
(1996a). That is Sample A, comprising 94 stars.  Sample B is an update and
extension of Sample A to 109 stars, where we have included the new observations
from this paper, and the improved abundances of Ryan et al. (1999), Norris et
al. (2000), and Spite et al. (2000), who give an abundance for CS~29527-015
which previously had only an upper limit on its Li abundance.  As in the
previous work, three least-squares regression routines have been employed: a
weighted-least-squares (WLS) approach; an unweighted (standard) least-squares
approach (LS); and a re-weighted least-squares approach based on the least
median of squares (RLS/LMS) algorithm with outlier rejection (see Ryan et al.
1996a and Appendix A of this paper for details).

Table~3 provides the coefficients of the fits and their standard 1$\sigma$
errors. The coefficient of determination, $R^2$, which is a measure of the
amount of variance in $A$(Li) that can be accounted for by the regression
model, is also given.  The main conclusion to be drawn from Table 3, seen by
comparing the Sample A and Sample B results, is that even though Ryan et al.
(1996a) were working with biased and inferior data to those now available, the
coefficients associated with $T_{\rm eff}$ and [Fe/H] are essentially unchanged
from the earlier analysis.  Addition of the hotter, more metal-rich stars has
not weakened any of the earlier arguments; in fact, the coefficients of
determination have all risen.  Of course, in and of itself, this does not
establish the {\it reality} of the $T_{\rm eff}$ trend, as we have
intentionally used the same temperature scale as before; the scale could
contain systematic errors as Bonifacio \& Molaro (1997) suggest in their IRFM
study.  We hope that the reality, or otherwise, of the $T_{\rm eff}$ trend will
be established reliably once the halo temperature scale is known with greater
certainty.  We note in closing this topic that the coefficients were not
significantly affected by choosing a higher low-temperature limit (5800~K) for
the sample.

In Figure 6(a) we plot the available halo data in the $A$(Li) vs [Fe/H] plane.
Observations of young- and old-disk stars by Boesgaard \& Tripicco (1986),
Rebolo, Molaro, \& Beckman (1988), Lambert, Heath, \& Edvardsson (1991), and
Nissen et al. (1999) are included to show the evolution of Li beyond the halo
phase.  The present sample of halo stars with [Fe/H]~$>~-2$ confirms that the
metallicity-dependence of $A$(Li) discussed above continues right up to the
highest halo metallicities.  However, as this figure contains a wide range of
effective temperatures, and we are concerned about a non-zero dependence of
$A$(Li) on that parameter (be it genuine or artificial), we restrict the sample
shown in Figure 6(b) to include only the hottest stars, those having with
$T_{\rm eff}~>$~6000~K.  The elimination of cooler stars, for which we derive
generally lower Li abundances, results in a narrower trend of $A$(Li) vs
[Fe/H].  In the following section we will use this sample to  constrain GCE
models in an attempt to develop a clearer picture of the evolution of this
element.

A recent analysis of BD+00$^\circ$2058A (King 1999) derives a considerably
higher Li abundance for this star, $A$(Li)~=~2.53$\pm$0.05, than we do, 
$A$(Li)~=~2.28$\pm$0.05 (1$\sigma$). These differ at the 3.5$\sigma$ level.
The King value implies considerably more Li GCE (at least for the material that
constitutes this object) than the measurements in our sample. Because of the
significance of this potentially high abundance for the remaining discussion,
we examine this difference in greater detail. The basic spectral measurements
of the star are in good agreement; we list $W$~=~37.6$\pm$3.3~m\AA, whereas
King measures $W$~=~42.3~m\AA\ from higher S/N and higher resolving-power data,
to which we would assign a measurement error of $\sigma_W$~=~1~m\AA. These
equivalent width measurements differ only at the 1.4$\sigma$ level, which is
quite reasonable, and lead to a difference in $A$(Li) of 0.05~dex. King infers
a temperature higher by 96~K --- a 1.1$\sigma$ difference --- which induces
another 0.06~dex abundance difference in the positive direction.  As described
previously, Li computations are largely insensitive to the gravity and
microturbulence, and we do not expect these differences in our analysis to lead
to significant abundance changes. The other major difference in the analyses is
the choice of model atmospheres. While our work (Ryan et al. 1996a) is based on
Bell (1983, priv.comm.) models that closely match older Kurucz (1989,
priv.comm.) models, King adopts the overshooting models of Kurucz (1993) which
are hotter in the shallower layers. The higher temperatures in these newer
models, even for an identical $T_{\rm eff}$, result in weaker lines being
computed, and hence higher abundances being inferred to match the observations.
This accounts for an additional difference of $\simeq$+0.11~dex (see Ryan et
al.  1996a, Fig.~2) for Li, and accounts at least partially for the higher
[Fe/H] derived by King; see Ryan et al. (1996b) for a discussion of model
differences in the context of elements other than Li, which show effects at a
similar order of magnitude.  Through the differences in observed line strength,
effective temperature, and choice of model, we are thus able to understand
0.22~dex of the +0.25~dex difference. In view of diminishing returns, we do not
endeavour to trace the remaining 0.03~dex difference. As all of the data
presented in our current work are on the same $T_{\rm eff}$ and model
atmosphere scale as those in the works by Ryan et al. (1996a,1999), we maintain
the data shown in our Table~1, without adjustment.  The Li abundance in
BD+00$^\circ$2058A is not, in our view, any more remarkable than the abundances
in the rest of the stars at that metallicity; King's (1999) impression to the
contrary emerged from the comparison of one star analysed on one system with
the bulk of data analysed on another.  We nevertheless acknowledge King's
superior data, and note that using King's equivalent width measurement on our
effective temperature and model atmosphere system would lead to
$A$(Li)~=~2.33$\pm$0.04.

Before concluding this section, we draw attention to the four stars in the
current survey for which no Li line was detectable.  Detection of four
extremely Li-deficient objects in a sample of 18 stars, when previous estimates
of the frequency of such objects in the halo population was ~5\% (Norris et al.
1997), is astonishing.  These objects are discussed in a companion paper (Ryan
et al.  2001) and will be excluded from further discussion in the present
work.

\section{Galactic Chemical Evolution of Li} %4

In this section, we compare the observed metallicity dependence of $A$(Li) with
several models for GCE, in the hope of gaining a better understanding of the
sources of this element. We explicitly assume that halo dwarfs with
[Fe/H]~$<~-1$ and $T_{\rm eff}~>$~6000~K have not depleted their pre-stellar
surface abundances {\it in-situ}.  However, the GCE models {\it do} allow for
astration, i.e., the removal of Li from the gas ``reservoir'' of star formation
via its destruction in stars that, at the end of their lives, re-mix with the
interstellar medium via winds and/or ejecta.  We also assume that the observed
abundances apply to pure $^7$Li only, i.e., that any pre-stellar $^6$Li has
been destroyed unless stated otherwise.  This could be incorrect for the
highest temperature stars in this parameter range, but is unlikely to
overestimate the $^7$Li abundance by more than 5\% (0.02 dex) based on the few
$^6$Li detections achieved.

\subsection {A Simple, Linear-Evolution ``Fiducial'' Model} %4.1

A simple analytic model could assume that post-primordial net production of Li
evolved linearly with iron, giving $n$(Li)~=~$n$(Li)$_{\rm p}$~+~$k\times
n$(Fe), where $A$(Li)$_{\rm p}$~ is the primordial abundance, and $k~=~{{{\rm
d}n{\rm (Li)}}\over{{\rm d}n{\rm (Fe)}}}$ is the relative rate of Li and Fe
nucleosynthesis (assumed in this simple model to be constant).  Two boundary
conditions fully specify the model: the primordial abundance of Li and the
meteoritic abundance. Adopting $A$(Li)$_{\rm p}$~=~2.10 for the former and
$A$(Li)$_{\rm m}$~=~3.30 for the latter (cf. Grevesse \& Sauval 1998), and
using $A$(Fe)$_\odot$~=~-4.50+12.00 yields $k~=~6.31\times 10^{-5}$, or one Li
atom produced (net) for every 16000 Fe atoms.

Such a linear-evolution model ignores all of the details of the physics of
nucleosynthesis and stellar life-cycles, and instead relies on an hypothesised
association of Li-producing events with Fe-producing events, be they due to
Galactic cosmic ray (GCR) spallation, the $\nu$-process, or some
other.\footnote{Parizot \& Drury (1999), for example, emphasise that the linear
relation between isotopes which emerges from their work is the result of
differential {\it dilution}, rather than accumulation. This model therefore
also follows a linear evolution path, and provides a very different example of
a complex physical model whose outcome can, with the benefit of hindsight, be
approximated over the halo epoch by a simple linear relation.} The model is
shown in Figure 7(a), where a remarkable similarity to the data can be seen.
The fact that such a simple model gets even close to the observations indicates
that the real process(es) responsible for Li production in the halo do indeed
follow Fe production almost linearly. However, the mismatch during the
evolution of the disk indicates that a more efficient source of Li production
relative to iron is required to reproduce the steep $A$(Li) vs [Fe/H] trend
exhibited by the disk data. We find the elements of such a model in the
work by Romano et al. (1999), as discussed below, in particular the ``late''
synthesis of Li in novae and AGB stars.

Changing the evolution assumption to linearity with the $\alpha$-elements,
where we adopt [$\alpha$/Fe]~=~0.4 for [Fe/H]~$<~-1$ and
[$\alpha$/Fe]~=~$-0.4\times$[Fe/H] otherwise, flattens the rate of Li evolution
in the disk (where $\alpha$-elements evolve more slowly with respect to iron;
this is expected on the basis of the time delay model of chemical evolution,
since Fe is mostly produced by SNeIa whereas alpha-elements are essentially
synthesized by SNeII).  As a consequence, the Li evolution in the halo phase
must be greater in order to reach the meteoritic abundance at [Fe/H]~=~0, with
the result that the model curve is higher under this scenario, and lithium is
clearly over-produced as compared to the observational data.

\subsection{The GCE Model of Fields \& Olive} %4.2

Fields \& Olive (1999a,b) developed a GCE model of post-primordial $^7$Li
production by GCR spallation and fusion (which also produce $^6$Li, $^9$Be, and
$^{10,11}$B), and by supernovae through the $\nu$-process (which also produces
$^{11}$B; Woosley et al. 1990). The GCR composition is assumed to scale with
the ISM composition, which leads to dominance of the fusion source (as opposed
to spallation) during halo-star formation (e.g., Steigman \& Walker 1992), and
linear evolution of the GCR Li contribution with the number of supernovae
during this era. The $\nu$-process likewise gives linear evolution with the
number of supernovae. The {\it normalisations} of these sources are set by the
meteoritic abundances; $^9$Be and $^{10}$B set the GCR contribution, and the
$^{11}$B unaccounted for by this means fixes the $\nu$-process contribution.

The number of Type II supernovae contributing to the production of Li in the
early Galaxy can, in principle, be traced by the abundances of heavier
elements.  Two possible tracers are oxygen and iron, the former made by
hydrostatic burning in the progenitor, the latter formed during the explosive
phase, and both expelled during the explosion. However, it is unclear which
element is a better tracer of supernova numbers; iron yields are notoriously
difficult to calculate due to its dependence on many factors associated with
the explosion (e.g., mass-cut, neutronization, rotation) (e.g., Timmes,
Woosley, \& Weaver 1995; Hix \& Thielemann 1996; Hoffman et al. 1999; Nakamura
et al.  1999), while observational studies of oxygen have often provided
inconsistent results (reviewed by Gratton 2000). The Fields \& Olive (1999a,b)
models use oxygen as the tracer, and utilise in particular the [O/Fe] results
of Israelian, Garc\'ia Lop\'ez, \& Rebolo (1998) and Boesgaard et al.
(1999).\footnote{We note that the issue of the correct value of the halo [O/Fe]
ratio remains contentious (e.g., Fulbright \& Kraft 1999).  The 2000 IAU
General Assembly has a one day joint discussion which hopefully will generate
more light than heat on this problem.}

The linearity of Li nucleosynthesis with the number of supernovae, both for the
GCR and $\nu$-process, partially justifies the choice of parameters in the
simple linear-evolution model described above. Fields \& Olive's model,
however, is based on proper physical processes, and is shown in Figure 7(b) for
$A$(Li)$_{p}$~=~2.04. Two curves are shown depending on whether the GCE $^6$Li
component is preserved or destroyed at the surface. This model was intended to
explore Li GCE in Population II stars only, and excludes additional stellar
production mechanisms that come into play during the evolution of the Galactic
disk. For this reason, the appropriate test of the model against the data
involves only the stars with [Fe/H]~$<~-1$, not the disk stars at higher
metallicity which the model does not address. It should also be noted that the
model shown was developed prior to the reduction and analysis of the new data
presented in this paper (for stars at [Fe/H]~$\sim~-1.5$), so the excellent fit
of the Fields \& Olive model to these new points is a genuine achievement of
the model.

Note that the predictions of the Fields \& Olive model, especially for the more
metal-rich stars, depend on the survival fraction of $^6$Li. Where $^6$Li has
been measured in stars with [Fe/H]~$\sim~-2.4$ (Smith, Lambert, \& Nissen 1993,
1998; Cayrel et al. 1999; Hobbs \& Thorburn 1994, 1997; Deliyannis \& Ryan
2000), its total fraction is low, but nevertheless consistent with the Fields
\& Olive model (see Ryan et al. 2000). Bear in mind also that $^6$Li retention
is likely to be a function of metallicity, as discussed by Ryan et al. (1999),
so the data might be expected to follow the dashed curve at lowest [Fe/H] and
the solid curve at higher [Fe/H].

\subsection{The GCE model of Romano et al. 1999} %4.3

The Li GCE model by Romano et al. (1999) includes five components: primordial
nucleosynthesis, GCR spallation (using the prescription of Lemoine,
Vangioni-Flam, \& Cass\'e 1998), supernova nucleosynthesis via the
$\nu$-process, AGB star nucleosynthesis via hot-bottom burning and the $^7$Be
transport mechanism (Cameron \& Fowler 1971; Sackmann \& Boothroyd 1992), and
novae (Jos\'e \& Hernanz 1998).  Of the post-primordial contributions, the
$\nu$-process dominated during the halo phase, so much so that Romano et al.
considered a model with the contributions from this process halved to avoid
over-production of Li.  AGB stars were found to contribute only for
[Fe/H]~$^>_\sim~-0.8$, and novae only for [Fe/H]~$^>_\sim~-0.5$.  Therefore,
for halo-star evolution only the primordial nucleosynthesis, the $\nu$-process,
and GCR processes are significant. These are the same contributions included in
the Fields' \& Olive model, albeit with different normalisations.  The
considerable contribution of novae and non-negligible contribution of AGB stars
at ``late'' times of GCE, i.e., for the disk, are promising candidates for the
sources required to raise $A$(Li) from the value of the most metal-poor disk
stars to the meteoritic one.

In Figure 7(c) we show Romano's model ``B'' plus their adopted GCR contribution
(based on the work of Lemoine et al. 1998).  Their model adopted
$A$(Li)$_p$~=~2.20, whereas our data suggest a lower value, so we also show a
second model with the primordial value reduced to $A$(Li)$_p$~=~2.10 to fit the
most metal-poor stars.  The Romano et al.  models provide a very good fit to
the disk-star data, and indicate that the substantial contribution of novae at
late times may indeed be the requirement to account for the steep evolution of
$A$(Li) in the disk.  The Romano et al.  models fit the most metal-poor halo
and disk data very well, but under-predict the abundances measured in the
present work for high-metallicity halo stars by $\simeq$0.08~dex.  
This difference could be due to a metallicity-dependent
systematic error in the color--effective-temperature calibration, if that
error changes by 100~K over the interval from [Fe/H] = -1.5 to
$-3.0$. An error of this magnitude is possible, but has not been identified.
If, on the other hand, the derived abundances
are reliable, then one would infer that the rate of Li
evolution in the halo phase of the model appears to be somewhat low,  
due to the use of the Lemoine et al. (1998) Li absolute yields from GCRs which would
then appear to underestimate GCR Li production during the halo phase.

\subsection{A Hybrid Model}

We found above that the Fields \& Olive model fit the halo evolution of the
Galaxy very well, while the Romano et al. model included the steep evolution
required during the disk phase. In this subsection, we examine a combination of
the two, using the primordial, GCR, and $\nu$-process calculations of Fields \&
Olive, and add the AGB star and novae contributions computed by Romano et al.
for the disk phase. This hybrid replaces the GCR prescription of Lemoine et al.
(1998) and the $\nu$-process yields of Woosley \& Weaver (1995) adopted by
Romano et al. with the ones of Fields \& Olive (1999a,b) described above.

We recognize that combining these model results is not self-consistent, in
that the full GCE calculation including astration was applied by Romano et al.
on the basis of the components {\it they} included, and not on the basis of the
Fields \& Olive components which we now seek to substitute. Clearly, a
self-consistent recalculation is desirable. However, astration appears to be a
fairly minor factor compared with the source terms, at least up to
[Fe/H]~=~$-0.6$, judging by the minor impact on the survival fraction of
primordial Li in the Matteucci et al. (1999, Fig.~9a) calculations and in the
Fields \& Olive model (Ryan et al.  2000, Fig. 1).  Consequently, we regard
our present hybrid approach as a valuable initial investigation of the effect
of combining these sources.

The hybrid model is shown in Figure 7(d).  The Fields \& Olive GCR + $\nu$
component is clearly weaker than that in Figure 7(c), because the hybrid model
does not reach the meteoritic value, but it does improve the fit to the halo
and old-disk data.  Although the young-disk data are not so well reproduced,
the appearance of nova and AGB star nucleosynthesis in the calculations of
Romano et al. coincides with the observed steepening of disk star Li evolution,
and the novae source term is regarded as quite uncertain. As emphasised by the
fiducial linear-evolution model (Figure 7(a)), Li must evolve considerably
faster relative to iron during formation of young-disk stars to reproduce the
data. Even though none of the models, on its own, produces a {\it perfect} fit
to {\it all} of the metal-poor halo, metal-rich halo, old-disk and young-disk
data, the key features of the hybrid of Romano et al. and Fields \& Olive
models are in very good accord with the observations, and strongly suggest that
these models are viable (though not necessarily uniquely so; see below).

\subsection{The Inhomogeneous GCE Model of Suzuki et al. 1999}

Several recent models of GCE propose that very metal-deficient stars were
formed in individual SN remnant (SNR) shells in the Galactic halo during its
early epochs before the gas of the ISM was well-mixed, and that their abundance
patterns reflect the contributions of elements  synthesized in {\it single}
SNIIe events (Ryan, Norris, \& Bessell 1991; Audouze \& Silk 1995; McWilliam et
al.  1995; Ryan, Norris, \& Beers 1996b).  Even for stars of similar iron
content, the observed abundances of several r-process elements in these stars
show remarkable scatter (Gilroy et al. 1988; Ryan et al. 1991, 1996b; McWilliam
et al. 1995).  This scatter is presumed to arise from the inhomogeneous
evolution of the early Galactic halo, in particular the finite extent of
supernova remnants responsible for early enrichment of the proto-Galactic gas
(Ryan et al. 1996b).  This framework casts doubt on the applicability of simple
one-zone models, such as those discussed in the previous subsections, for
describing the chemical evolution of the Galaxy. Lithium must also be included
in any paradigm shift of this sort.  Tsujimoto et al. (1999; also Argast et al.
2000) have proposed a SN-induced chemical evolution model which takes account
of inhomogeneous circumstances arising from the stochastic nature of star
formation processes triggered by SN explosions. Their model can explain the
scatter seen in the Eu abundance, and can be applied to other elements,
e.g. iron, as well.

An extension of the inhomogeneous model to investigate the evolution of the
light elements Be and B, which are mainly produced by nuclear reactions
involving GCRs, has been developed by Suzuki et al. (1999).  They proposed a
new scenario, that GCRs originate from both the SN-ejecta and the swept-up ISM
accelerated by the shock formed in the SN remnant shell, and demonstrated that
this model reproduces the observed trends of Be and B very well.  The Galactic
cosmic rays accelerated by SN shocks propagate through the inhomogeneous
Galactic halo to interact with both the ambient ISM and the gas in SNR shells,
producing Be and B.  Their model exhibits a linear increase of log(BeB)
$\propto$ [Fe/H] quite naturally.  They suggested, for the first time, that
there might be expected a good correlation between time since the initiation of
star formation in the early Galaxy and $^6$LiBeB abundances, even for
low-metallicity stars [Fe/H]~$\le~-2$, an epoch when no unique time-metallicity
relation is expected to exist for heavier elements that are more affected by
the inhomogeneous nature of the early Galactic halo (Suzuki et al.  2000b;
Beers, Yoshii \& Suzuki 2000).  The implications for the chemical evolution of
Li should therefore be investigated in this inhomogeneous GCE model.

In the inhomogeneous model presented here, contributions from five components
are included, as in the model of Romano et al. (1999):  primordial
nucleosynthesis, GCR spallative and $\alpha+\alpha$ fusion reactions, SN
nucleosynthesis via the $\nu$-process (Woosley \& Weaver 1995), AGB star
nucleosynthesis (Forestini \& Charbonnel 1997), and nova nucleosynthesis
(Jos\'e \& Hernanz 1998).  The predicted $^{7}$Li yields of Woosley \& Weaver
(1995) for different SN progenitor masses are used for the $\nu$-process, but
the absolute values are reduced by a factor of five in order to match the
observed $^{11}$B/$^{10}$B ratio.  Note that the same model setup is applied to
all light elements $^{6,7}$Li, $^9$Be and $^{10,11}$B without any adjustable
parameter for each element.  To extend their model smoothly to later times, 
evolution of the disk ([Fe/H] $>$ --1.0) assumes a simple one-zone chemical
evolution model with infall (e.g., Pagel 1997, Chapter 8).

Figure~7(e) displays the result obtained in our model calculation.  The
primordial lithium abundance is chosen to be $A({\rm Li})_p=2.09$, derived from
likelihood analysis (Suzuki, Yoshii, \& Beers 2000a) by comparing the
theoretical frequency distribution of stars with recent accurate observations
by Ryan et al. (1999).  The frequency distribution is convolved with a Gaussian
having $\sigma=0.03$ dex for $A({\rm Li})$ and $\sigma=0.15$ dex for [Fe/H], in
order to compare with the observed data directly.  The long-dashed contours
indicate the probability densities of the predicted stellar distribution of
$^7$Li in the Galactic halo.  The solid curve in the center represents the
average trend of the evolution of $^7$Li vs Fe; the short-dashed curve gives
the total $^6$Li~+~$^7$Li for when $^6$Li is preserved.  The model accounts for
the observed halo data very well.  Note that a clear departure of Li from the
primordial abundance, at [Fe/H]~$^<_\sim~-3$,  towards higher metallicity,
[Fe/H]~$\sim~-1$, is apparent, supporting the contention that the observed
gradient with [Fe/H] is real.  The source of the low-abundance slope is mainly
due to Li GCR production, at first ([Fe/H]~$<~-2$) from the $\alpha+\alpha$
fusion reactions,  and later also from spallation reactions as the CNO
abundance in the ISM increases.  The contribution from the $\nu$-process of SNe
is less than 10\% of the total post-primordial Li production for the halo
phase.  Since the predicted increasing trend of Li is greater in stars with
metallicity $-2~\le$~[Fe/H]~$\le~-1$ than for more metal-poor stars, additional
observations of turnoff dwarfs in this metallicity region would help
measure the gradient more precisely.  They are also needed to identify
uncertain contributions from AGB and novae nucleosynthesis in the old disk
phase, where the model slightly over-produces the data.

\subsection{Discussion of GCE Models of Li, and Future Directions}

We note at the outset of this discussion that, at first sight, there are few
{\it major} differences between the evolutionary trends in the models presented
above.  The models shown in Figure 7(a), (c), and (e), for example, have quite
different underlying assumptions, but predict very similar evolution of Li
during the halo phase, especially when the error bars on the observations are
taken into account. The reason is that these three models all have mechanisms
which are linear with Fe; the underlying assumptions and parameter sets are
different, but the proportionality is the same. This highlights a degeneracy in
attempts to find the ``best'' model by comparisons with imperfect data; models
based on quite different propositions give rise to very similar GCE histories.
The major differences between models arise during the disk phase, which is also
where Li evolution is greatest and Li abundances are more easily measured.
However, due to the considerable changes in the sources and sinks of Li that
operate throughout the course of Galactic history, one must not be fooled into
discarding a particular model for halo evolution based on the failure of its
disk component, for example. Some of the more subtle differences between the
models have already been discussed, but we note the following before
concluding.

Of the models presented above, those in Figure 7(c), 7(d) and 7(e) include both
halo and disk sources. None provides a perfect match to both Galactic
components. All, of course, are ``fixed'' to the data for the most metal-poor
halo stars through the choice of the primordial value.  The Romano et al. model
provides a good match to the metal-poor disk and meteoritic values, but does
not exhibit sufficient Li evolution during the halo phase, and over-predicts
the Li abundance at [Fe/H]~$\simeq~-0.5$; the observations show a later onset
for high Li-production. The hybrid model, which combines the halo evolution of
Fields \& Olive with the Romano et al.  disk results, provides a better fit to
the metal-rich halo and metal-poor disk data up to and including the stage
[Fe/H]~$\simeq~-0.5$, but does not provide sufficient Li production during the
later stages of the disk, and fails to reach the meteoritic value. The
inhomogeneous model of Suzuki et al.  likewise produces a good fit to the halo
data within a quite different framework, but the one-zone disk calculation to
which it is connected is inadequate. As the mismatch between the disk
observations and simple fiducial model emphasises, the efficiency of Li
production relative to iron must increase substantially during the late phase
of disk evolution if the Li enhancement at [Fe/H]~$>~-0.5$ is to be reproduced.

The models have emphasised the importance of the GCR contribution to Li during
halo evolution. The Fields \& Olive (1999a,b) and Suzuki et al. (1999) models
have both assumed a traditional energy spectrum for the $\sim$100--1000~MeV
cosmic rays, but one should also be aware that a quite different class of
energetic particles may play an important role. Several authors have
examined the possible role of a shock-accelerated low-energy component (LEC)
which dominates the particle flux at a few~$\times$~10~MeV, the threshold for
Li production (e.g., Ramaty, Kozlovsky, \& Lingenfelter 1996; Vangioni-Flam et
al. 1998).  Ramaty et al. and Lemoine, Vangioni-Flam, \& Cass\'e (1998; also
adopted by Romano et al. 1999), for example, conclude that the solar/meteoritic
light-element abundances can be best fit with contributions from both the GCR
and LEC components in ratios 1:3 or 1:1 respectively.  Moreover, Vangioni-Flam
et al. (1998,1999) show that the LEC can dominate light-element production
during the halo phase, the traditional GCR contribution becoming significant
only during evolution of the disk. A better understanding of this component
is essential to obtaining correct models of Li evolution throughout
Galactic history. As the Vangioni-Flam et al. (1998) LEC models invoke
acceleration in superbubbles (e.g., Parizot \& Drury 1999), treatment of this
mechanism in an {\it inhomogeneous} halo environment (a la Suzuki et al. 1999)
would also be a valuable undertaking.

Another remarkable feature of the models shown in Figure~7 is that the range we see
in $A$(Li) values for new observations at [Fe/H] $>$ $-2$ is
consistent with the modelled evolutionary rates of Li, and does not require a
spread in Li about the trend in excess of that due to measurement errors alone.
The one exception amongst our new observations in CD-30$^o$18140, which sits
above the curves. It will be interesting to see whether future investigations 
of this
object, including a detailed stellar atmosphere analysis,
confirms it as lying above the curves.

Note that we have avoided any discussion of {\it time}-rates of evolution; the
mismatch with the fiducial model emphasises that Li production relative to {\it
iron} must increase. The challenge for astronomers is to identify the source of
that Li.  The models discussed above have so far excluded a recently recognised
source of Li, namely cool-bottom processing in low-mass red giants (e.g.,
Sackmann \& Boothroyd 1999). This has been used to explain the observations of
Li-rich stars in this phase of evolution (e.g., Charbonnel \& Balachandran
2000; de la Reza, da Silva, \& Drake 2000; Gregorio-Hetem, Castilho, \& Barbuy
2000).\footnote{The high Li abundance found in the C-rich star CS~22898-027
(Thorburn \& Beers 1992) may reflect material transferred from a giant
companion. However, the presence of s-process enhancements in this particular
object (McWilliam et al. 1995) suggests contamination from an AGB rather than
RGB former primary.} The contributions of these stars to the overall chemical
evolution of the Galaxy is not yet known, but it is clear by their low mass
that they will contribute only late in the evolution of the system. We might
speculate, therefore, that such objects could be significant contributors to
disk evolution that will provide the required higher efficiency of Li
production relative to iron at [Fe/H]~$>~-0.5$.

\section{Conclusions} %5

We have measured the Li abundances of 18 stars with $-2~<$~[Fe/H]~$<~-1$ and 
6000~K~$<~T_{\rm eff}~<$~6400~K, populating a previously poorly-sampled region
of the $T_{\rm eff}$, [Fe/H] plane. Of these, four proved to be highly
Li-deficient, with $A$(Li)~$<~1.7$ (and are discussed in the companion paper).
The remaining 14 were found to conform to the same trends of $A$(Li) with
$T_{\rm eff}$ and [Fe/H] identified earlier by Ryan et al. (1996a), removing
any doubts that the results of that study were affected by the selection biases
inherited in the observational samples.

It is unclear whether the $T_{\rm eff}$ trend we have identified is intrinsic
or due to the photometric effective temperature scales used, but this
uncertainty was largely circumvented by restricting our attention to stars in a
narrow $T_{\rm eff}$ range by excluding those with $T_{\rm eff}~<~6000$~K.
This sub-sample of turnoff halo stars, supplemented with objects from previous
studies occupying the same temperature range, revealed a {\it significant
increase} of $A$(Li) over the metallicity range of the halo, from
$A$(Li)~=~2.10 at [Fe/H]~=~$-3.5$ to $A$(Li)~=~2.40 at [Fe/H]~=~$-1.0$.

We examined various GCE models in an attempt to understand the halo and disk
phases of Li production, and showed, with a simple linear-evolution model, that
the net Li production rate relative to iron must increase substantially during
young-disk evolution.  A very satisfactory match to the halo and old-disk data
was found in the three-component (primordial, GCR, and $\nu$-process) model of
Fields \& Olive (1999a,b; Ryan et al. 2000). The additional sources of stellar
nucleosynthesis in the young disk ([Fe/H]~$>~-0.5$)  are well represented by
the models of Romano et al. (1999), whose main contributors are novae, and to a
lesser extent, AGB stars. A hybrid of the two models provided the best current
match to the halo and disk data together. In addition, a new model of halo GCE
in an inhomogeneous framework, extending the work of Suzuki et al. (1999), was
presented which was equally capable of modelling the halo data.

While none of these models presents a {\it perfect} fit to {\it all} epochs of
Galactic evolution, the match between the models and data is sufficiently good
to believe that the models are viable, albeit not uniquely so.  The primary
remaining challenge is to reproduce the efficient production of Li during late
stages of disk evolution. A simple fiducial model demonstrates that Li
production relative to iron production increases significantly at
[Fe/H]~$>~-0.5$. The disk model of Romano et al. currently comes closest to
predicting this, but we also speculate that the recently identified cool-bottom
processing (production) of Li in low-mass red giants (Sackmann \& Boothroyd
1999) may provide a late-appearing source of Li without attendant Fe
production.

\section{Acknowledgements}

The authors gratefully acknowledge the support for this project given by the
Australian Time Assignment Committee (ATAC) and Panel for the Allocation of
Telescope Time (PATT) of the AAT and WHT respectively, and for practical
support given by the staff of these facilities.  T.C.B. acknowledges partial
support from NSF grant AST 95-29454, and along with S.G.R., wishes to thank the
IAU for travel supplement grants which enabled them to attend IAU Symposium
198, where discussions related to this work were held.

\appendix

\section{Appendix}

The existence, or not, of correlations between the estimated abundance,
$A$(Li), and the physical parameters $T_{\rm eff}$ and [Fe/H], individually or
in a multiple regression approach, has been the subject of a number of recent
papers.  For transparency, we include our complete regression results for the
samples discussed in this paper.

\subsection{Data Samples and Methodology}

The data sets we consider are:

\noindent Sample A:  The values of $A$(Li), $T_{\rm eff}$, and [Fe/H]
published by Ryan et al. (1996a). 
(The regressions published by Ryan et al. (1996a) were 
based on a penultimate version of the data, prior to final updates for a few
stars. We have also changed our approach for carrying out the weighted least 
squares analysis as discussed below, so our present values, though differing 
little, supersede those reported previously).  

\noindent Sample B:  Our estimates of $A$(Li), $T_{\rm eff}$, and [Fe/H] 
presented in this paper, plus the data from Ryan et al. (1996a), 
Ryan et al. (1999), Norris et al. (2000), and Spite et al. (2000).
Where a star appears twice, the most recent data have been used.
Stars with only upper limits on $A$(Li) have been excluded.

\noindent Sample C:  The sample of ``metal-rich'' stars shown in Figure 5.

The models we consider are the following:

\noindent Model 1:  $A$(Li) = $A_0 + A_1\;T_{\rm eff}/100$

\noindent Model 2:  $A$(Li) = $A_0 + A_2\; {\rm [Fe/H]}$

\noindent Model 3:  $A$(Li) = $A_0 + A_1 T_{\rm eff}/100 + A_2\; {\rm [Fe/H]}$

The regression approaches we utilize are the following:

\noindent Technique 1:  WLS -- A weighted least-squares approach, wherein the
regression is weighted by taking into account the reported statistical error in
$A$(Li).  In Ryan et al. (1996a) we made use of the routines published in 
Bevington (1969).
In the present application, we choose to employ a different set of routines,
those given in the program SYSTAT 9.0, and described in Wilkinson, Blank, \&
Gruber (1996).  These approaches, and hence their results,  are slightly
different, in that the Bevington routines obtain predicted errors on the
regression coefficients which are, in general, a factor of 2--3 lower than
those reported by SYSTAT 9.0 (although the derived coefficients are essentially
identical).  This occurs because the error estimates from Bevington routines do
not explicitly take into account the residuals of the points from the derived
regression lines, but rather, assume that the statistical errors fully reflect
the expected level of error, which may not be the case for such a diverse
data set.

\noindent Technique 2:  LS -- A standard least-squares approach, which does not
take into account the statistical errors on $A$(Li), as obtained by SYSTAT 9.0.

\noindent Technique 3:  RLS/LMS -- A reweighted least-squares approach based on
the least median of squares method of Rousseeuw \& Leroy (1987).  This
approach implements an objective identification of outliers based on deviations
from a resistant regression fit (using LMS) obtained from multiple resamples of
the data.  Once identified, these outliers are removed and a standard
(unweighted) least-squares method is applied to the surviving data.  This
technique makes use of the code provided by Dallal (1991).

\subsection {Regression Results}

The results for these regressions are summarized in Table A1.  Column (1)
identifies the sample under consideration.  Column (2) lists the numbers of
stars in each sample.  Column (3) provides the regression model which is
reported. Column (4) lists alternative cuts on effective temperature.  Column
(5) indicates the regression technique which is applied.  Columns (7)--(9)
lists the derived regression coefficients and their one-sigma standard errors. 
Column (10) is the coefficient of determination, $R^2$, which quantifies the
amount of variation of $A$(Li) that can be accounted for by the regression
model under consideration (note, $0 \le R^2 \le 1$).  Column (11) lists, for
the bivariate regressions (Model 3), the Pearson correlation coefficients
between the independent variables $T_{\rm eff}$ and [Fe/H] obtained by the
technique under consideration, and is one indication of the presence of
possible collinearity between the predictor variables. 

We consider the results for each of our samples in turn.  

\subsection {Sample A}

As was the case in Ryan et al. (1996a), we identify a significant correlation 
with effective
temperature (considered in isolation -- Model 1)  for the case where stars of
the full range of $T_{\rm eff}$'s are considered.  The significance of the
correlation coefficient $A_1$ decreases, as expected, when more aggressive cuts
on the lower limit of effective temperatures are made.  However, it is
illuminating that the size of the coefficient remains roughly constant, on the
order of $A_1 \sim 0.03-0.035$ dex per 100~K for the different temperature
regimes.  This suggests that the presence of a temperature-related correlation
is not crucially dependent on the inclusion or exclusion of $A$(Li) estimates
for stars near the lower limits, where concerns about the possible depletion of
surface Li abundance are presently thought to have their greatest effect.

When metallicity is considered in isolation (Model 2), we find that the
correlation coefficient $A_2$ is small, and not significant, for the full
Sample A, and the subset of Sample A with $T_{cut} > 5800$ K.  However, all
three of the regression techniques return a significant correlation of $A$(Li)
with [Fe/H], roughly $A_2 = 0.10$ dex per dex, when the subsample of
stars with $T_{\rm eff} > 6000$K is considered. This is an indication that
a bivariate fit is required.

For the case of the bivariate regression model (Model 3), we note that, as in
Ryan et al. (1996a), significant coefficients on both $T_{\rm eff}$ and [Fe/H] are
returned when the full range of temperatures is considered, with a slightly
decreasing significance as more aggressive cuts on temperature are considered.
It is interesting to note that, for all temperature cuts, the coefficient of
determination indicates that between 25\% and 55\% of the observed variation in
$A$(Li) in the sample can be accounted for by the regression models, which is
much higher than seen for the case of either variable considered in isolation.

\subsection {Sample B}

The application of the above approaches to our expanded and refined sample 
of stars show some interesting differences as compared to the published sample
of Ryan et al. (1996a).  

For Model 1, the significance of the coefficient on temperature, $A_1$, drops
somewhat compared Sample A when
stars of all temperatures are included.  For the cuts in temperature,
$T_{\rm eff} > 6000$~K and $T_{\rm eff} > 5800$~K, $A_1$ becomes 
non-significant, a fact also
revealed from inspection of the coefficients of determination.  Nevertheless,
the value of the {\it significant} correlation, obtained for the subsample
including stars of all temperatures, is of the same order of magnitude, $A_1
\sim 0.03$ dex per 100K, as was found for the Ryan et al. (1996a) data set.  

For Model 2, the significance of the coefficient on [Fe/H], $A_2$, is markedly
higher than obtained from the application of this model to Sample A above, even
when the temperature cuts are applied. Similarly, the coefficients of
determination are higher as well.  This reflects the fact that Sample B
includes stars of a wider range in abundance than Sample A, and also, that the
errors in estimated $A$(Li) have been substantially reduced for a number of the
lowest metallicity stars from the work of Ryan et al. (1999).  
The value of $A_2$ increases
from $A_2 \sim 0.07$ dex per dex to $A_2 \sim 0.12$ dex per dex as one considers
progressively more aggressive cuts on temperature.  For stars with 
$T_{\rm eff} > 6000$~K, the coefficient of determination rises to 
$R^2 \sim 0.5-0.6$,
indicating that more than 50\% of the variation in $A$(Li) is accounted for by
this model.

For Model 3, which we consider the most appropriate, the significance of both
of the coefficients $A_1$ and $A_2$ remain high, and the coefficients of
determination are clearly much higher than obtained for Sample A above.  The
correlation coefficients between the $T_{\rm eff}$ and [Fe/H] variables have
decreased somewhat for the subsample that includes stars of all temperatures,
but are roughly similar to those obtained previously for the two temperature
cuts. This result indicates that we are making progress with respect to
reducing the possible influence of collinearity in the predictor variables, but
that further work remains -- a doubling or tripling of the numbers of stars
with available Li measurements in the metallicity range $-2 \le {\rm [Fe/H} \le
-1$ would be most helpful.  

\subsection{Sample C}

For the ``metal-rich'' stars which comprise this sample, the regression
coefficient on temperature obtained for Model 1 is both larger, on the order of
$A_1 \sim 0.05-0.07$ dex per 100K, and markedly more significant (note the
associated dramatic rise in the values of the coefficients of determination)
than was found for either Sample A or Sample B considered above.  The opposite
statement can be made concerning the coefficients on [Fe/H], $A_2$, which have
decreased to non-significance in this subsample.  This is perhaps not
surprising, as the exclusion of the lower values of [Fe/H] 
should be expected to have a significant effect on the derived correlations.
Interestingly, this result also suggests that previous considerations of this
problem, going back to the original claim of Spite \& Spite (1982), might have
been unduly influenced by the lack of available measurements of $A$(Li) for
stars of the lowest metallicity.  

When the two predictors are considered in a bivariate regression model, such as
Model 3, the significance of $A_1$ remains high, while that of $A_2$ increases
(at least for the temperature cut $T_{\rm eff} > 5800$), reaching marginal ($3-4
\sigma$) significance.  The coefficients of determination are also
significantly increased in the bivariate regression model, as compared to
Models 1 and 2.  It should be noted that, by excluding the most metal-deficient
stars, the collinearity of the predictor variables is markedly decreased.

% \vfill
% \eject
% {\bf Figure captions}

\clearpage
\begin{figure}[!htb]
\begin{center}
\leavevmode
\epsfxsize=160mm
% \epsfysize=60mm
% \epsfbox[40 185 570 305]{cmd.ps}
\epsfbox{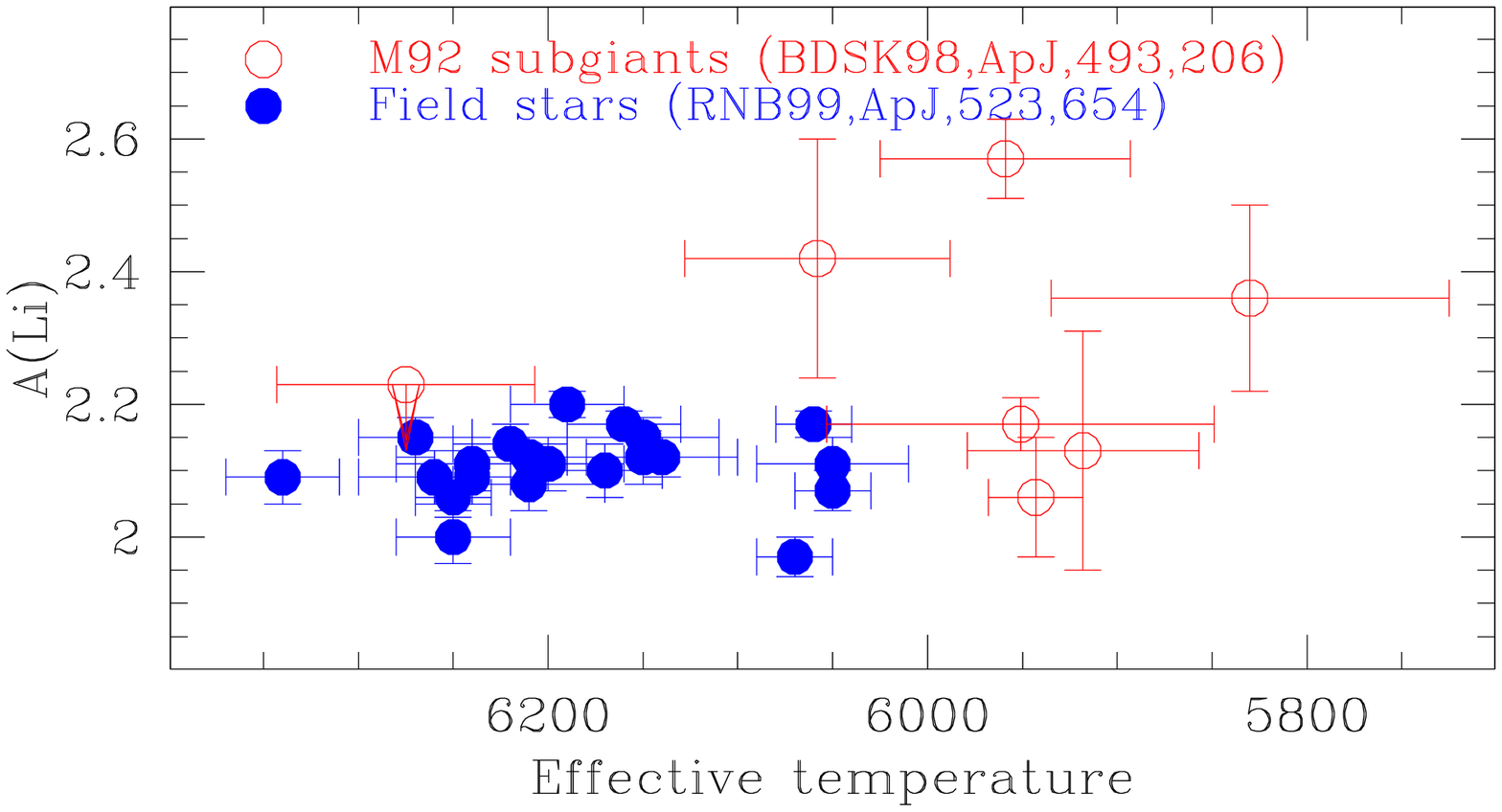}
\end{center}
% % psfile=#1 vsize=#2 angle=#3 hscale=#4 vscale=#5 hoffset=#6 voffset=#7
% \plotfiddle{sample_fig1.eps}{75truemm}{0}{45}{45}{-150}{-80}
\caption{
Fig. 1: 
Li spread in halo field and globular cluster samples.
}
\end{figure}

\clearpage
\begin{figure}[!htb]
\begin{center}
\leavevmode
\epsfxsize=160mm
% \epsfysize=60cm
% \epsfbox[40 185 570 305]{cmd.ps}
\epsfbox{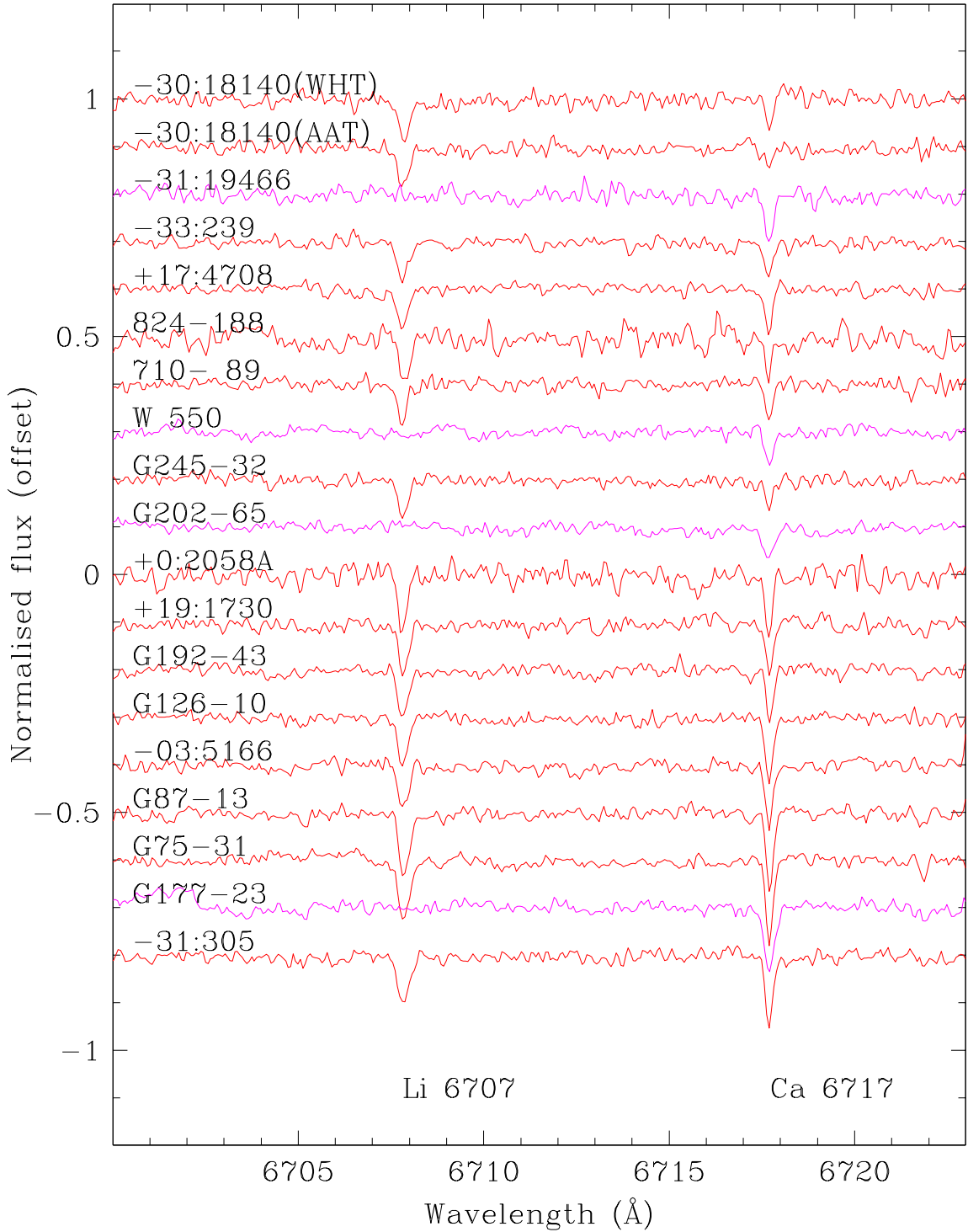}
\end{center}
% % psfile=#1 vsize=#2 angle=#3 hscale=#4 vscale=#5 hoffset=#6 voffset=#7
% \plotfiddle{sample_fig1.eps}{75truemm}{0}{45}{45}{-150}{-80}
\caption{
Fig. 2: 
Spectra in region of Li~6707 doublet, in order of increasing [Fe/H]. 
Note the presence of four stars with greatly depressed Li abundances.
{\bf NOTE TO EDITOR: TWO-COLUMN WIDTH FOR FIGURE 2 PLEASE.}
}
\end{figure}

\clearpage
\begin{figure}[!htb]
\begin{center}
\leavevmode
\epsfxsize=160mm
% \epsfysize=60cm
% \epsfbox[40 185 570 305]{cmd.ps}
\epsfbox{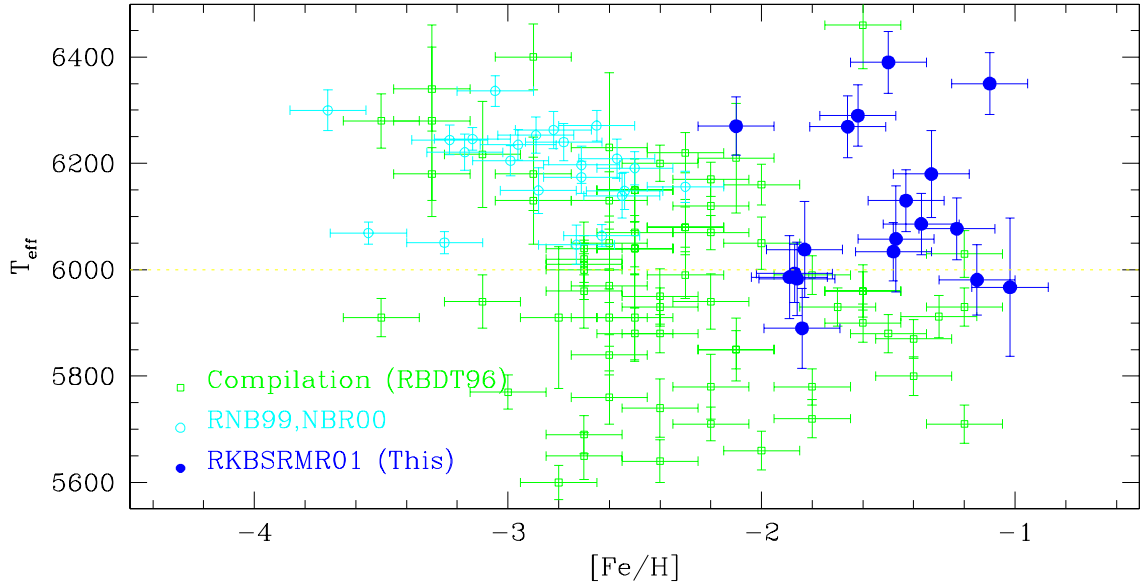}
\end{center}
% % psfile=#1 vsize=#2 angle=#3 hscale=#4 vscale=#5 hoffset=#6 voffset=#7
% \plotfiddle{sample_fig1.eps}{75truemm}{0}{45}{45}{-150}{-80}
\caption{
Fig. 3:
Location of program stars in the $T_{\rm eff}$ vs [Fe/H] plane, where they fill
a deficit caused by selection biases in the literature. {\it Dotted line:}
dividing line in sample at $T_{\rm eff}$~=~6000~K.
}
\end{figure}

\clearpage
\begin{figure}[!htb]
\begin{center}
\leavevmode
\epsfxsize=160mm
% \epsfysize=60cm
% \epsfbox[40 185 570 305]{cmd.ps}
\epsfbox{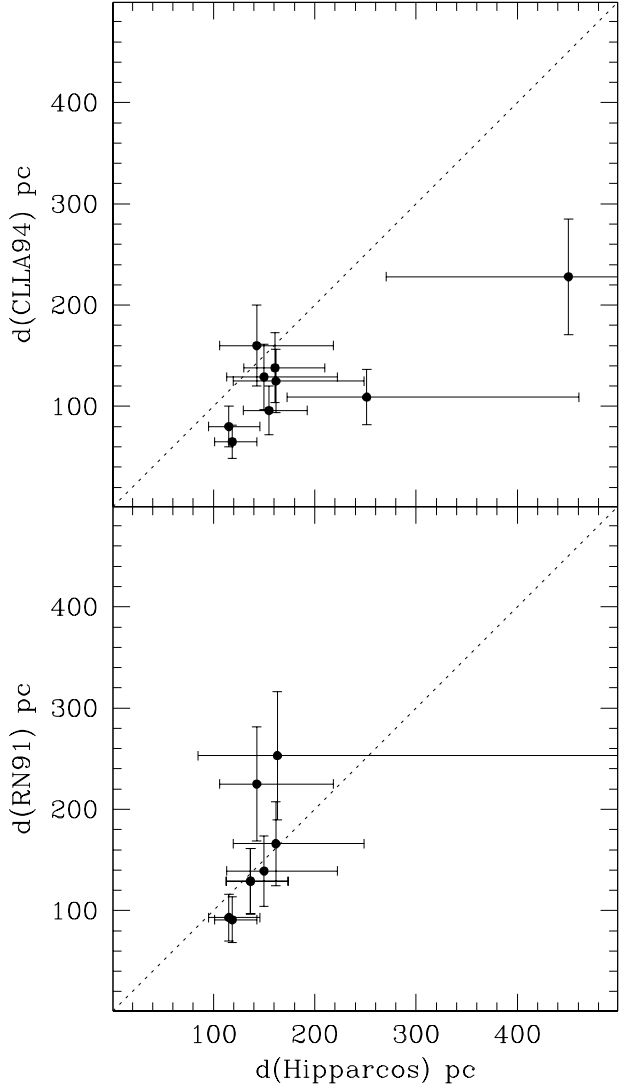}
\end{center}
% % psfile=#1 vsize=#2 angle=#3 hscale=#4 vscale=#5 hoffset=#6 voffset=#7
% \plotfiddle{sample_fig1.eps}{75truemm}{0}{45}{45}{-150}{-80}
\caption{
Figure 4:
Comparison of photometric distance scales of Carney et al. (1994) and
Ryan \& Norris (1991) with Hipparcos data. Uncertainties in the
photometric scales are taken as 25\%, whereas the Hipparcos uncertainties
are taken directly from the catalog.
}
\end{figure}

\clearpage
\begin{figure}[!htb]
\begin{center}
\leavevmode
\epsfxsize=160mm
% \epsfysize=60cm
% \epsfbox[40 185 570 305]{cmd.ps}
\epsfbox{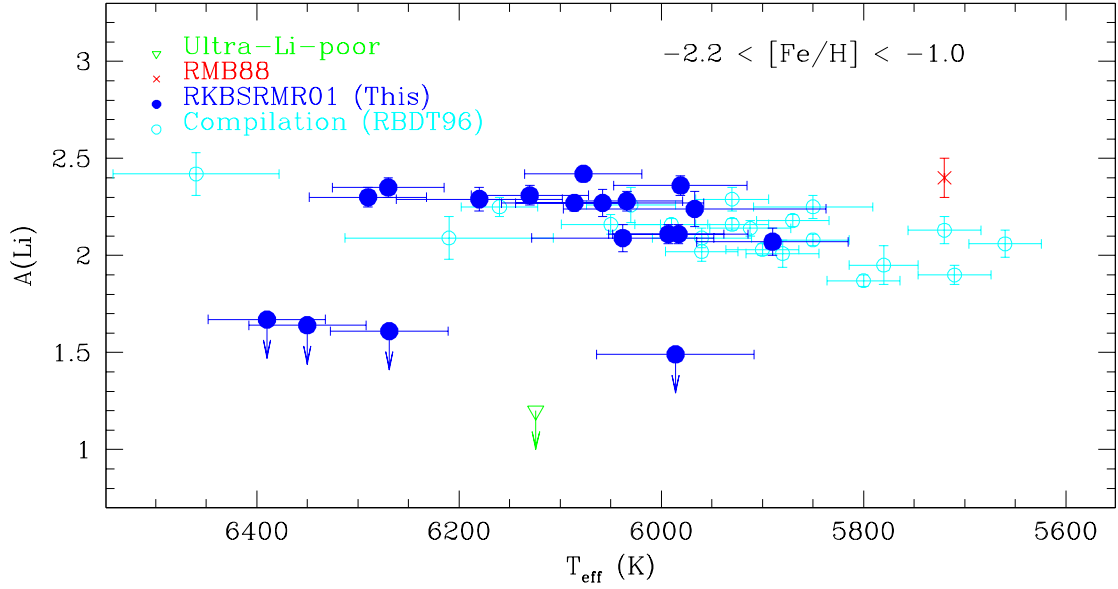}
\end{center}
% % psfile=#1 vsize=#2 angle=#3 hscale=#4 vscale=#5 hoffset=#6 voffset=#7
% \plotfiddle{sample_fig1.eps}{75truemm}{0}{45}{45}{-150}{-80}
\caption{
Fig. 5:
Variation of $A$(Li) with $T_{\rm eff}$ for stars with 
$-2.2~<~$[Fe/H]~$<~-1.0$. {\it Solid symbols}: new data; {\it open symbols}:
previous data from literature. It is unclear whether the trend is real or is
caused by deficiencies in the color-temperature transformation, but the new
data indicate that it is not due to selection biases in previous studies.
Note also the four ultra-Li-depleted stars.
}
\end{figure}

\clearpage
\begin{figure}[!htb]
\begin{center}
\leavevmode
\epsfxsize=160mm
% \epsfysize=60cm
% \epsfbox[40 185 570 305]{balmer.ps}
\epsfbox{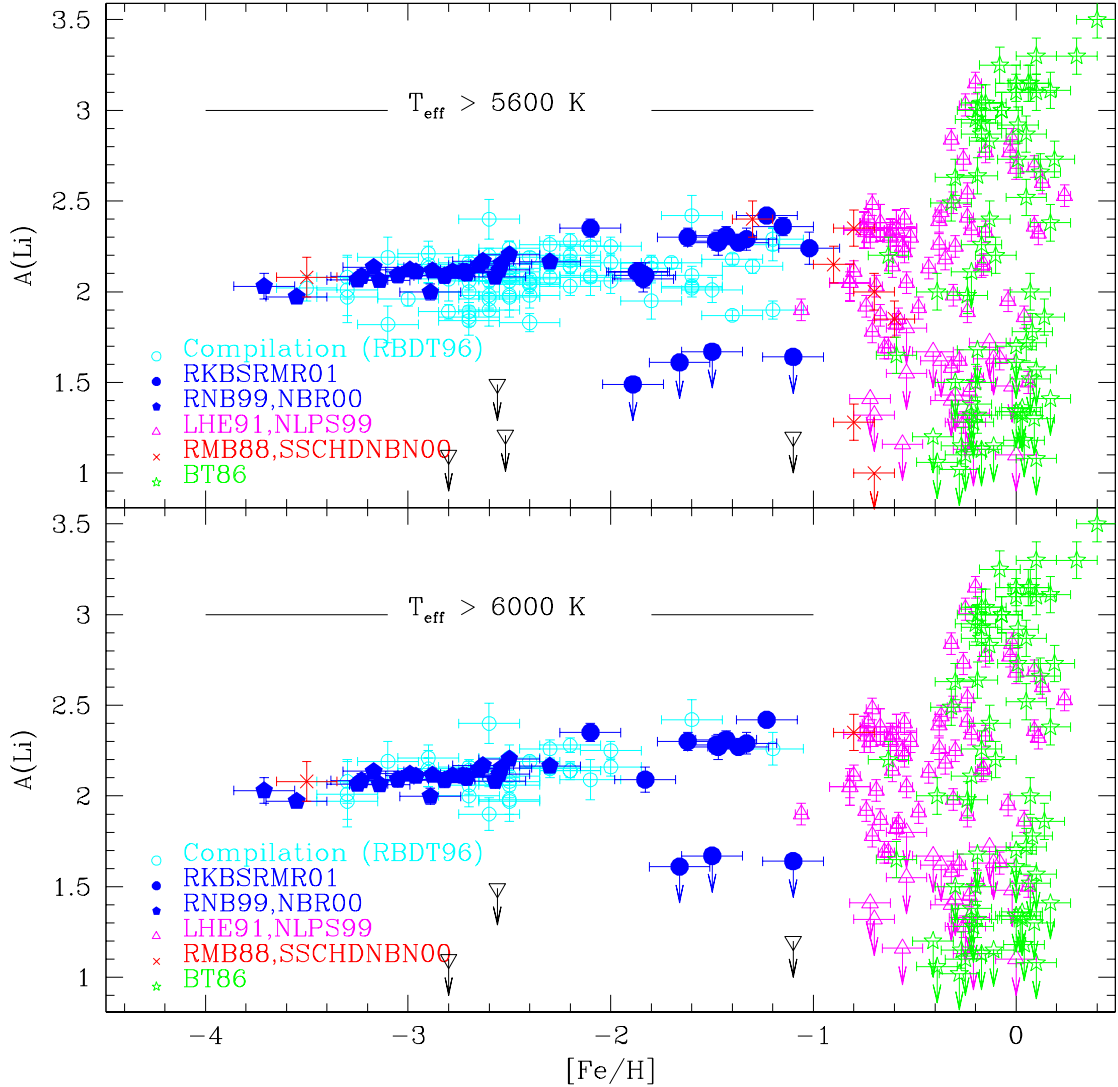}
\end{center}
% % psfile=#1 vsize=#2 angle=#3 hscale=#4 vscale=#5 hoffset=#6 voffset=#7
% \plotfiddle{sample_fig1.eps}{75truemm}{0}{45}{45}{-150}{-80}
\caption{
Fig. 6:
(a) $A$(Li) on [Fe/H] for halo, young disk and old disk. 
Data are from 
Boesgaard \& Tripicco (1986), Rebolo et al. (1988), 
Lambert et al. (1991), Nissen et al. (1999), Spite et al. (2000),
the homogenised compilation by Ryan et al. (1996a), and recent
data by Ryan et al. (1999) and Norris, Beers, \& Ryan (2000). 
The halo stars have $T_{\rm eff}~>$~5600~K.
(b) As for (a), but restricting the halo sample to $T_{\rm eff}~>$~6000~K 
to avoid the (genuine or artificial) $T_{\rm eff}$-dependence.
}
\end{figure}

\clearpage
\begin{figure}[!htb]
\begin{center}
\leavevmode
\epsfxsize=120mm
% \epsfysize=60cm
% \epsfbox[40 185 570 305]{balmer.ps}
\epsfbox{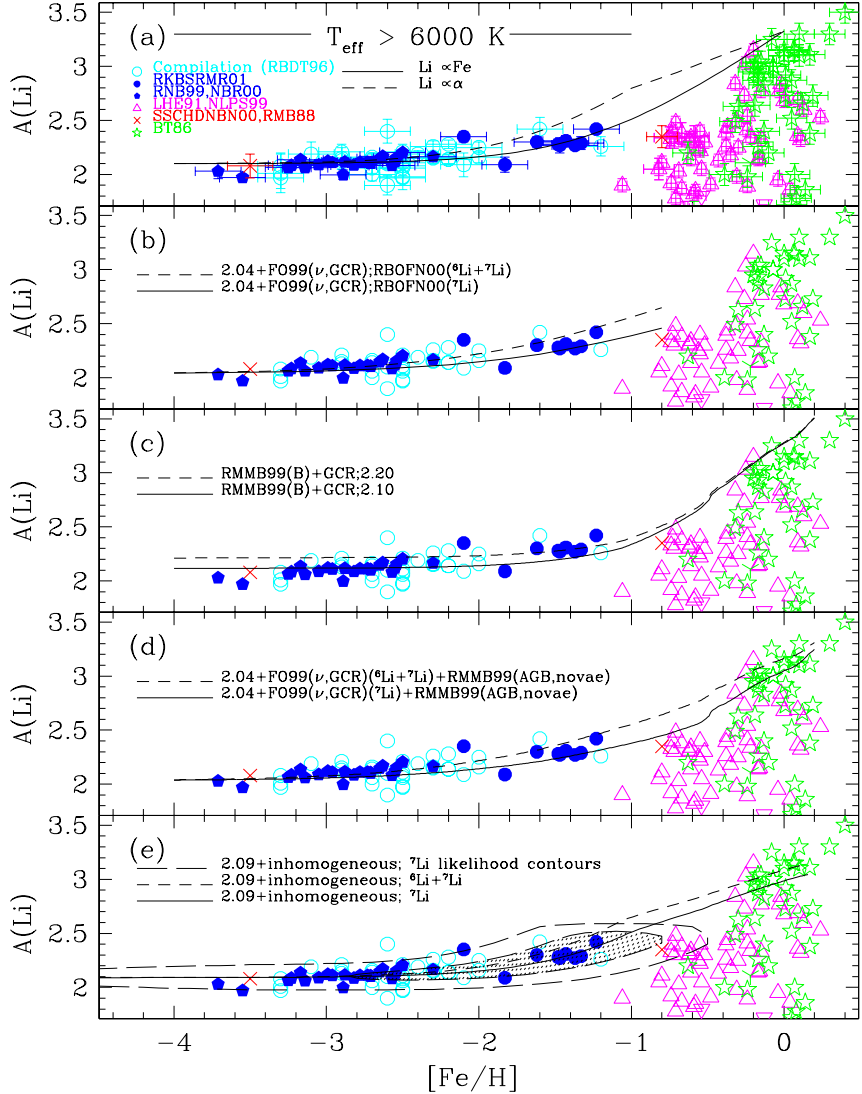}
\end{center}
% % psfile=#1 vsize=#2 angle=#3 hscale=#4 vscale=#5 hoffset=#6 voffset=#7
% \plotfiddle{sample_fig1.eps}{75truemm}{0}{45}{45}{-150}{-80}
\caption{
Fig. 7:
Comparison of halo, old disk and young disk star observations (Figure~6) 
with theoretical models. 
(a) Simple two-component model. {\it Solid curve}: assumes Li production scales
with iron; {\it dashed curve}: assumes Li production scales with the 
$\alpha$-elements.
(b) Primordial, GCR, and $\nu$-process model of 
Fields \& Olive (1999a,b; Ryan et al. 2000). {\it Solid curve}: $^7$Li only;
{\it dashed curve}: includes $^6$Li.
(c) Five component --- primordial, GCR, $\nu$-process, AGB star, and novae
model of Romano et al. (1999). {\it Solid curve}: adopts $A$(Li)$_p$~=~2.10;
{\it dashed curve}: $A$(Li)$_p$~=~2.20.
(d) Hybrid model using (b) plus the AGB star and novae contributions from (c).
{\it Solid curve}: $^7$Li only; {\it dashed curve}: includes $^6$Li.
(e) Inhomogeneous model. The two long-dashed contour lines, from the inside 
outwards, correspond to the (error-convolved) frequency distribution of 
long-lived stars of
constant probability density $10^{-4}$ and $10^{-8}$ in unit
area of $\Delta$[Fe/H]=0.1$\times\Delta A$(Li)=0.002. 
(The inner contour is shaded for clarity.)
{\it Solid curve}: evolution of the $^7$Li gas abundance.
{\it Short-dashed curve}: $^6$Li~+~$^7$Li abundance.
{\bf NOTE TO EDITOR: TWO-COLUMN WIDTH FOR FIGURE 7 PLEASE.}
}
\end{figure}

\clearpage
\begin{figure}[!htb]
\begin{center}
\leavevmode
\epsfxsize=175mm
% \epsfysize=60cm
% \epsfbox[40 185 570 305]{balmer.ps}
\epsfbox{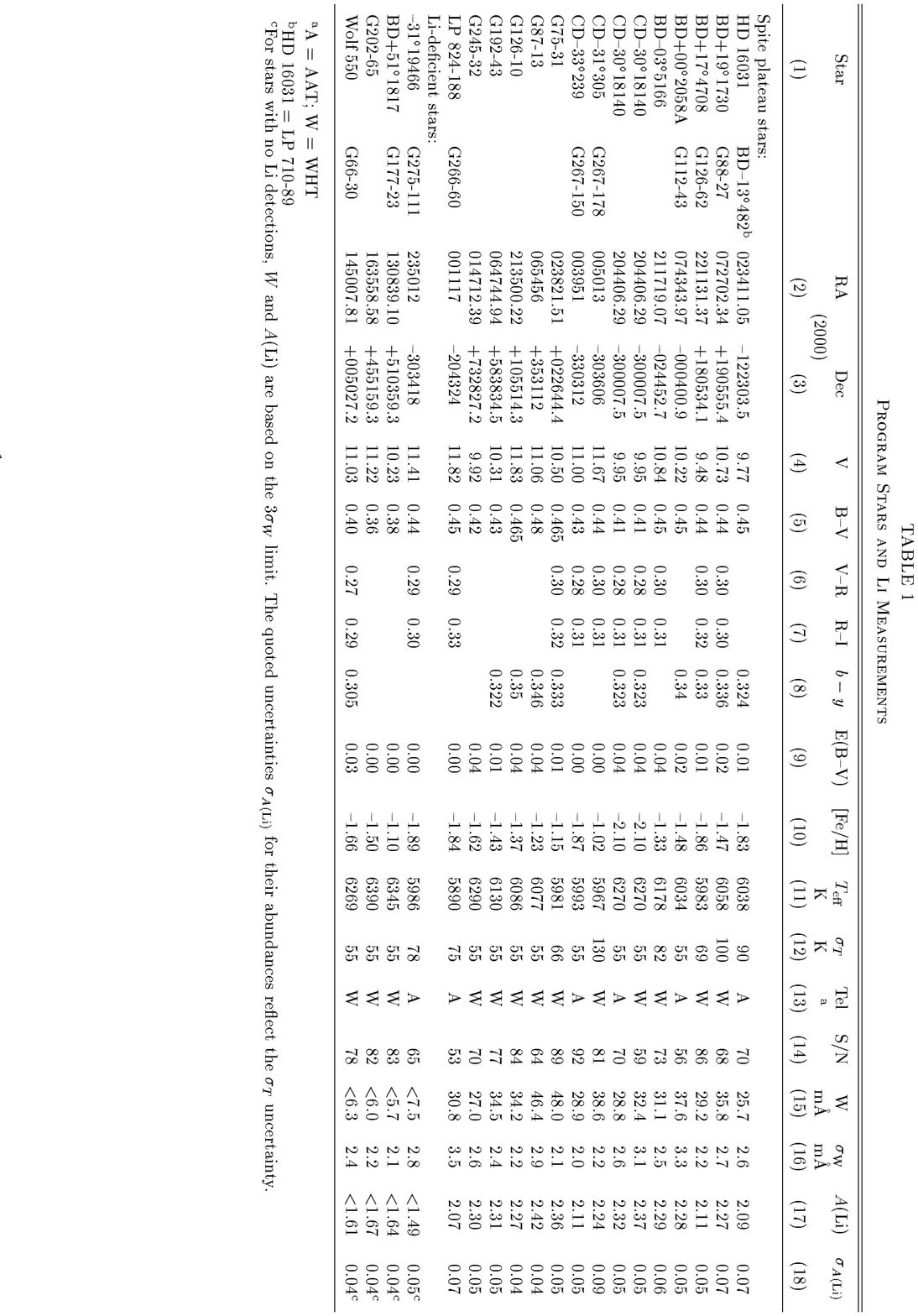}
\end{center}
% % psfile=#1 vsize=#2 angle=#3 hscale=#4 vscale=#5 hoffset=#6 voffset=#7
% \plotfiddle{sample_fig1.eps}{75truemm}{0}{45}{45}{-150}{-80}
\caption{Table 1}
\end{figure}

\clearpage
\begin{figure}[!htb]
\begin{center}
\leavevmode
\epsfxsize=175mm
\epsfbox{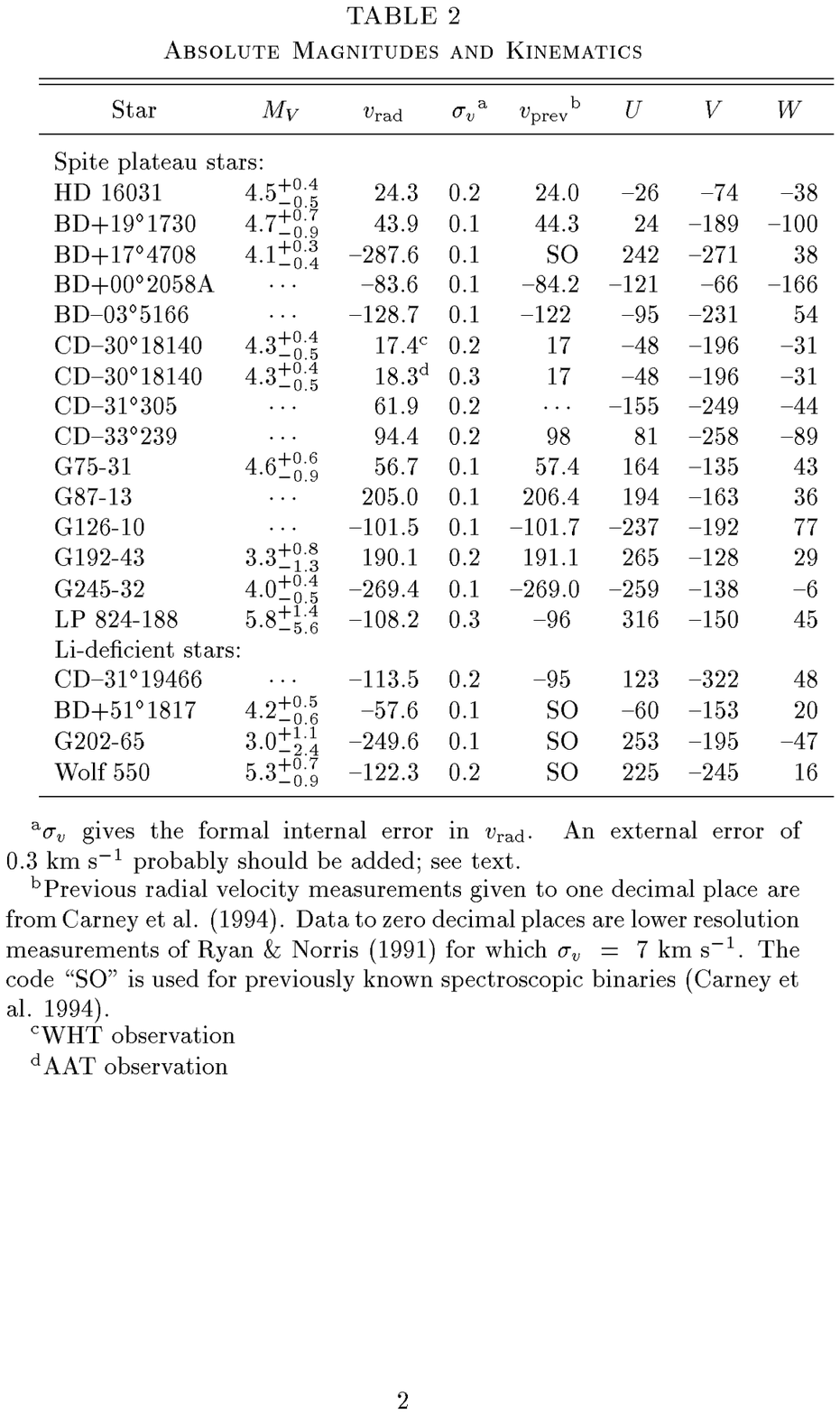}
\end{center}
% % psfile=#1 vsize=#2 angle=#3 hscale=#4 vscale=#5 hoffset=#6 voffset=#7
% \plotfiddle{sample_fig1.eps}{75truemm}{0}{45}{45}{-150}{-80}
\caption{Table 2}
\end{figure}

\clearpage
\begin{figure}[!htb]
\begin{center}
\leavevmode
\epsfxsize=175mm
\epsfbox{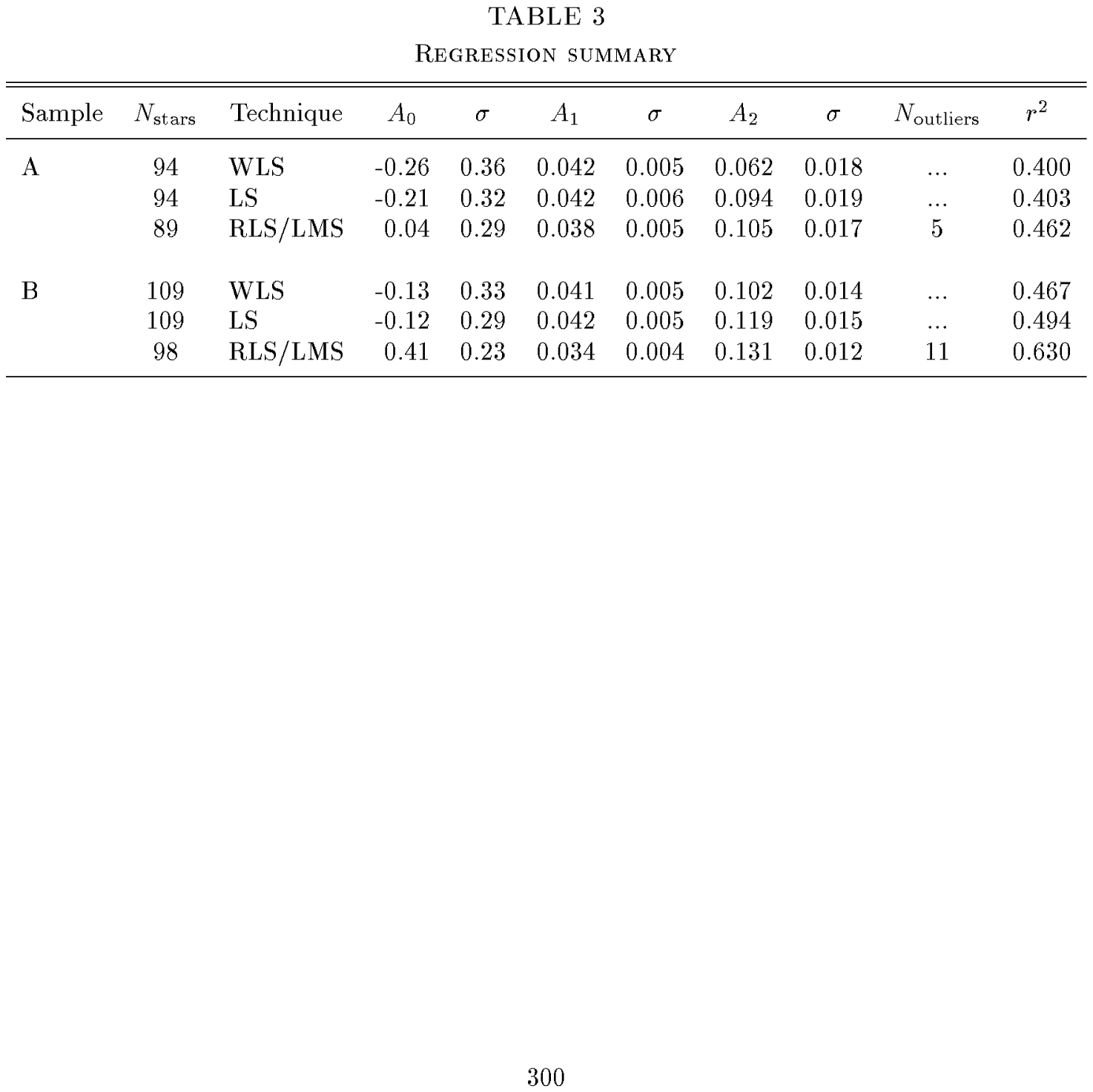}
\end{center}
% % psfile=#1 vsize=#2 angle=#3 hscale=#4 vscale=#5 hoffset=#6 voffset=#7
% \plotfiddle{sample_fig1.eps}{75truemm}{0}{45}{45}{-150}{-80}
\caption{Table 3}
\end{figure}

\clearpage
\begin{figure}[!htb]
\begin{center}
\leavevmode
\epsfxsize=175mm
\epsfbox{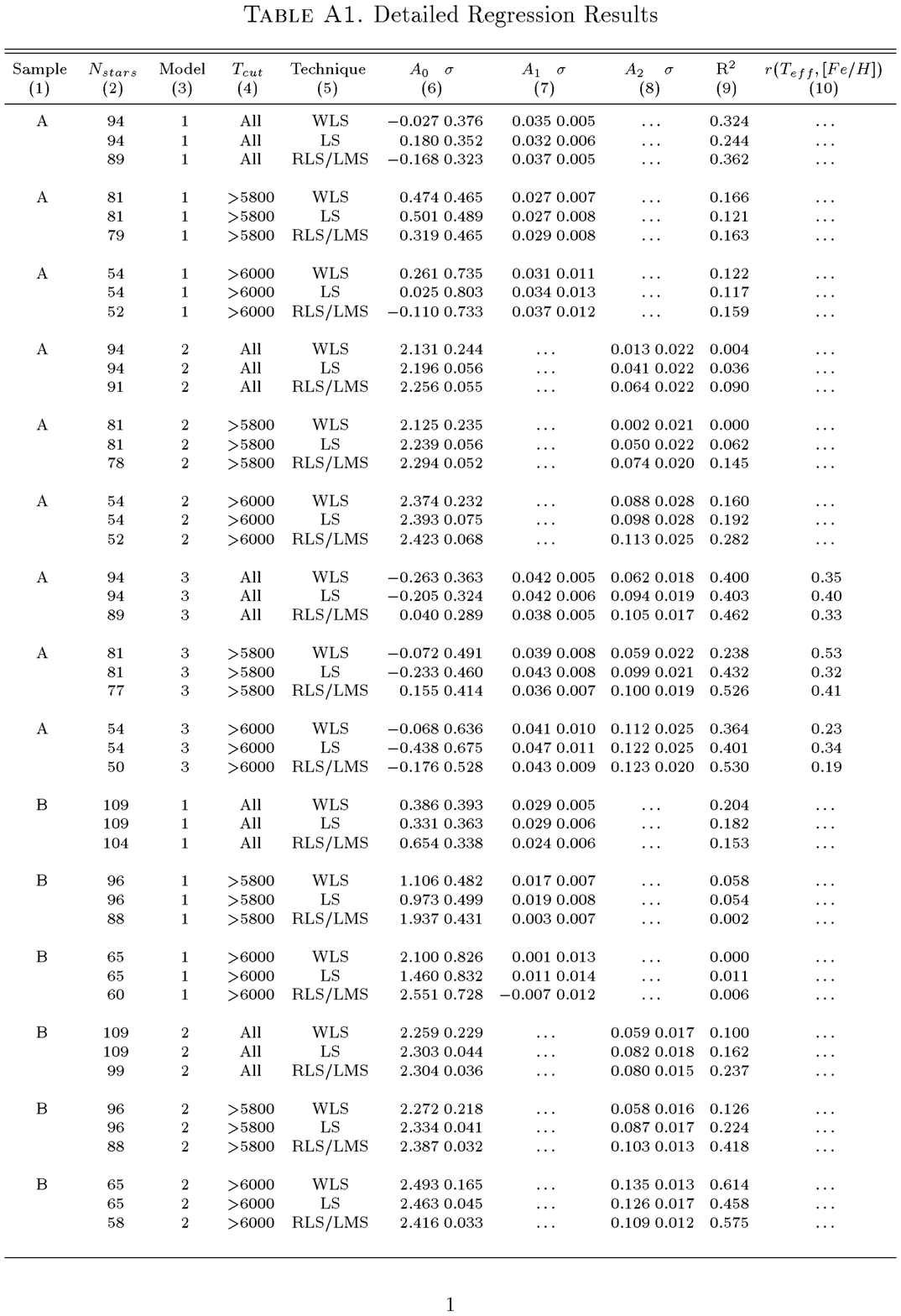}
\end{center}
% % psfile=#1 vsize=#2 angle=#3 hscale=#4 vscale=#5 hoffset=#6 voffset=#7
% \plotfiddle{sample_fig1.eps}{75truemm}{0}{45}{45}{-150}{-80}
\caption{Table A1(1)}
\end{figure}

\clearpage
\begin{figure}[!htb]
\begin{center}
\leavevmode
\epsfxsize=175mm
\epsfbox{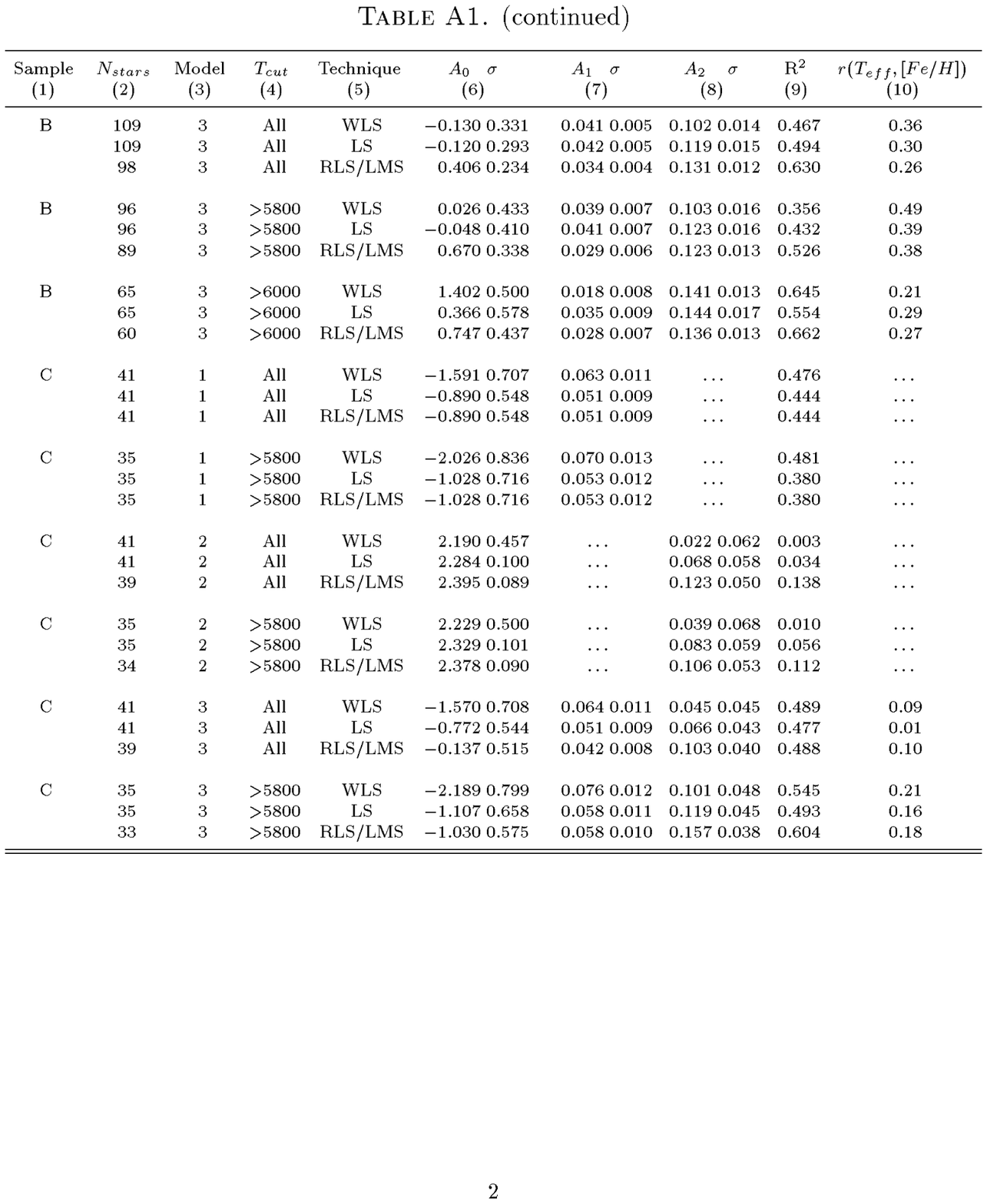}
\end{center}
% % psfile=#1 vsize=#2 angle=#3 hscale=#4 vscale=#5 hoffset=#6 voffset=#7
% \plotfiddle{sample_fig1.eps}{75truemm}{0}{45}{45}{-150}{-80}
\caption{Table A1(2)}
\end{figure}

\end{document}